\documentclass[a4paper,12pt]{article}
\usepackage[top=2.5cm, bottom=2.5cm, left=2.5cm, right=2.5cm]{geometry}
\usepackage{authblk, graphicx, amsmath, amsfonts, booktabs, verbatim, multicol, multirow, float, lineno}
\usepackage[hidelinks]{hyperref}

\title{Luminosity Performance of the \\ Compact Linear Collider at 380 GeV \\ with Static and Dynamic Imperfections} 
\author[1, 2]{C. Gohil}
\author[1]{P.\,N. Burrows}
\author[2]{N. Blaskovic Kraljevic\footnote{Present address: ESS, Lund, Sweden.}}
\author[2]{A. Latina}
\author[2]{J. \"{O}gren}
\author[2]{D. Schulte}
\affil[1]{\small John Adams Institute (JAI), University of Oxford, Oxford, OX1 3RH, United Kingdom}
\affil[2]{\small The European Organization for Nuclear Research (CERN), Geneva 23, CH-1211, Switzerland}
\date{\small (\today)}

\begin{document}
\maketitle

\begin{abstract}
The Compact Linear Collider is one of the two main European options for a collider in a post Large Hadron Collider era. This is a linear $e^+e^-$ collider with three centre-of-mass energy stages: 380 GeV, 1.5 TeV and 3 TeV. The luminosity performance of the first stage at 380 GeV is presented including the impact of static and dynamic imperfections. These calculations are performed with fully realistic tracking simulations from the exit of the damping rings to the interaction point and including beam-beam effects in the collisions. A luminosity of $4.3\times10^{34}\,\text{cm}^{-2}\text{s}^{-1}$ can be achieved with a perfect collider, which is almost three times the nominal luminosity target of $1.5\times10^{34}\,\text{cm}^{-2}\text{s}^{-1}$. In simulations with static imperfections, a luminosity of $2.35\times10^{34}\,\text{cm}^{-2}\text{s}^{-1}$ or greater is achieved by 90\% of randomly misaligned colliders. Expressed as a percentage of the nominal luminosity target, this is a surplus of approximately 57\%. Including the impact of ground motion, a luminosity surplus of 53\% or greater can be expected for 90\% of colliders. The average expected luminosity is $2.8\times10^{34}~\text{cm}^{-2}\text{s}^{-1}$, which is almost twice the nominal luminosity target.
\end{abstract}

\section{Introduction}
One of the top priorities of the 2013 European Strategy for Particle Physics Update~\cite{strategy2013} was to perform R\&D for a high-energy $e^+ e^-$ collider in Europe. The Compact Linear Collider (CLIC) is one of the two main options for an $e^+ e^-$ collider in Europe. Recently, several reports were submitted to the 2018 European Strategy for Particle Physics Update, describing the accelerator complex~\cite{clic-accelerator} and physics potential~\cite{clic-physics, clic-summary} of this collider. This paper reports on the luminosity performance of this collider.

\subsection{CLIC}
CLIC~\cite{clic-accelerator, clic-pip, clic-cdr} is a TeV-scale linear $e^+ e^-$ collider under development by the CLIC Collaboration. CLIC incorporates a staged approach with three centre-of-mass energies: 380\,GeV, 1.5\,TeV and 3\,TeV. The first stage at 380\,GeV, which is the focus of this paper, has been optimised for studies of the Higgs boson and top-quark physics~\cite{clic-higgs, clic-top}. 

The integrated luminosity goal for the 380\,GeV stage of CLIC is 180\,fb$^{-1}$ per year~\cite{clic-pip}. Assuming 185 days of operation and 75\% availability~\cite{clic-pip}, this corresponds to a nominal luminosity of
\begin{equation}
\mathcal{L} = 1.5\times 10^{34}~\text{cm}^{-2}\text{s}^{-1}.
\label{e:nominal-luminosity-target}
\end{equation}

\begin{figure}[!htb]
\centering
\includegraphics[width=\linewidth]{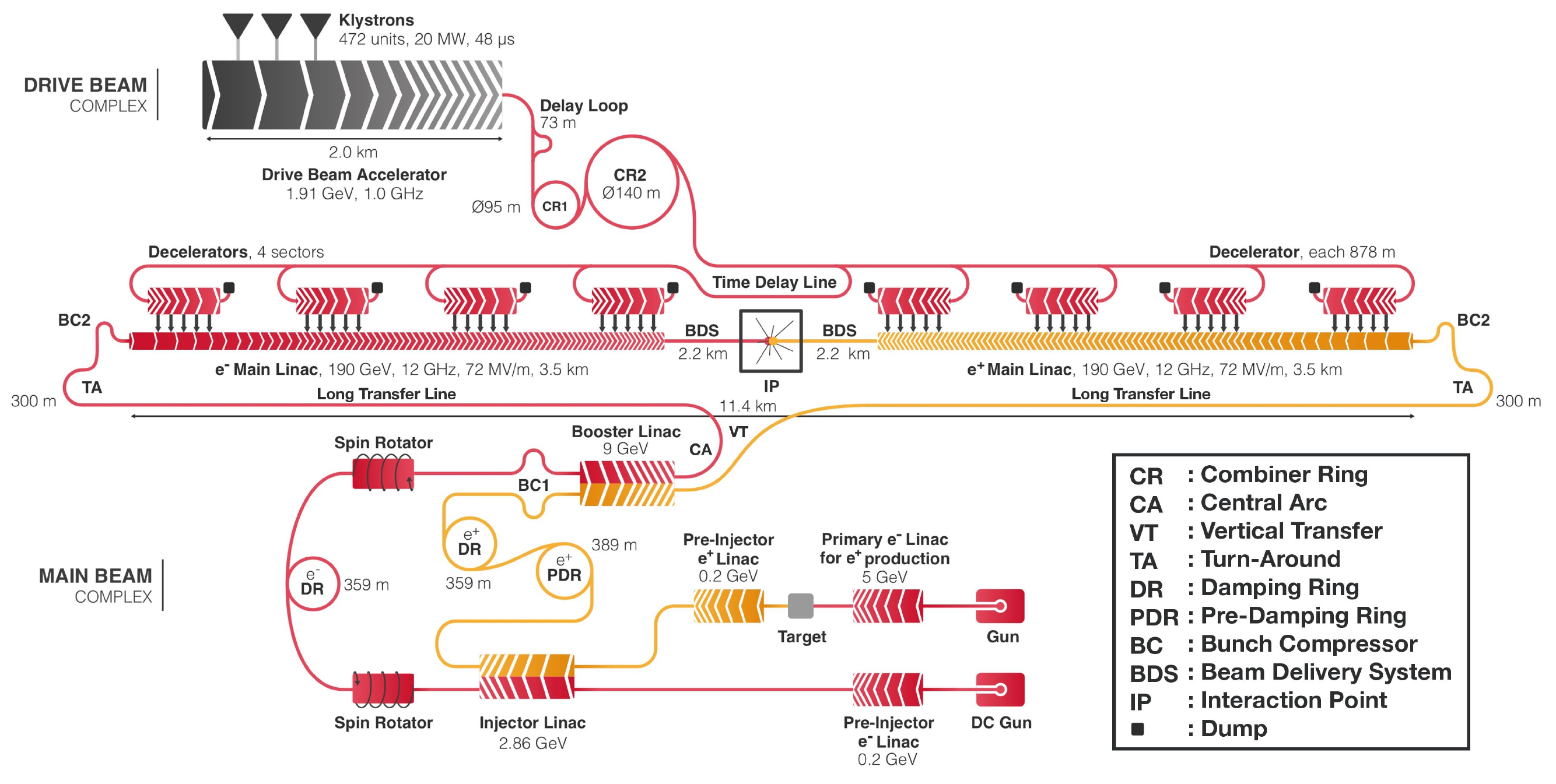}
\caption{\small Schematic diagram of CLIC at 380 GeV.}
\label{f:clic-380-gev}
\end{figure}

The 380\,GeV stage of CLIC is described in detail in~\cite{clic-pip}. A schematic of the beamline is shown in Fig.\,\ref{f:clic-380-gev}. The baseline is the drive-beam-based design with the Beam Delivery System (BDS) described in~\cite{fabien}.

CLIC utilises a novel two-beam acceleration scheme~\cite{clic-cdr}. This scheme involves using the power from a high-current, low-energy drive beam to accelerate a low-current main beam to high energies. Each beam has its own accelerator complex. In this paper, we study the main beam.

\begin{figure}[!htb]
\centering
\includegraphics[width=0.6\linewidth]{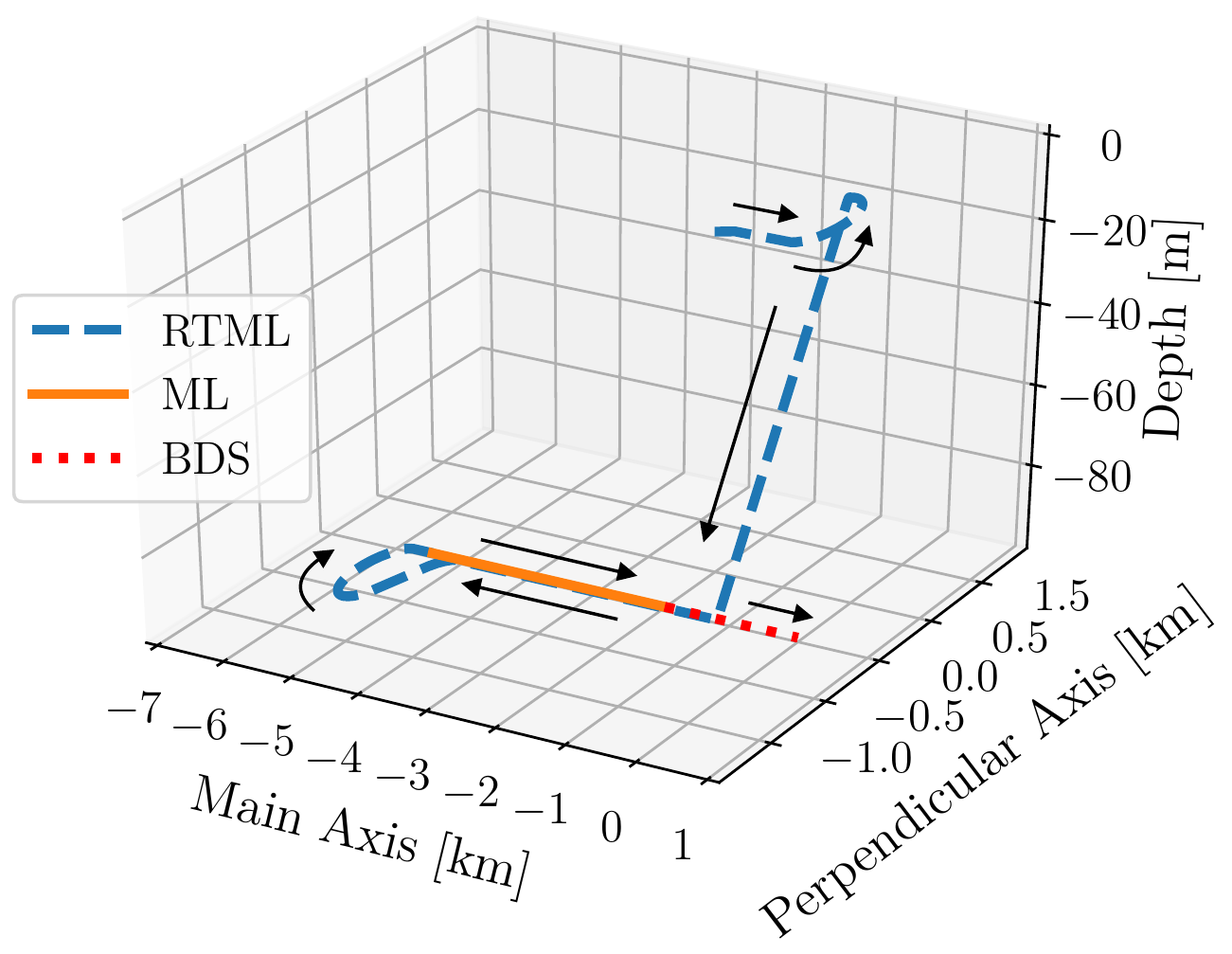}
\caption{\small Geometry of the electron beamline of CLIC at 380\,GeV. Main and perpendicular axis vs depth: RTML (dashed blue), ML (solid orange) and BDS (dotted red). The black arrows show the direction of the beam. The IP is at (0,0,-100).}
\label{f:clic-geometry}
\end{figure}

The main beam is transported from the Damping Ring (DR) to the Interaction Point (IP) through three sections: the Ring to Main Linac (RTML), Main Linac (ML) and BDS. The RTML contains all the sub-systems between the DR and ML shown in Fig.\,\ref{f:clic-380-gev}. The geometry of the electron beamline is shown in Fig.\,\ref{f:clic-geometry}. The beam is generated on the surface and is transported 100\,m underground in the RTML.

\subsection{Previous Studies}
Imperfections in beamline elements degrade the luminosity of a collider. Simulation studies are performed to determine the impact of imperfections. The ability of CLIC to reach its luminosity target in the 3\,TeV stage has been studied in detail in~\cite{clic-cdr}. In this paper, we study the luminosity performance of the 380\,GeV stage.

Beam dynamics studies for CLIC at 380\,GeV are summarised in~\cite{clic-pip}. Most efforts have focused on tuning studies of individual sections with static imperfections, namely the RTML~\cite{rtml-tuning}, ML~\cite{ml-tuning} and BDS~\cite{bds-tuning}. Previous studies of dynamic imperfections in CLIC at 380 GeV are limited to ground motion in individual sections. The impact of ATL motion in the ML and BDS is studied in~\cite{neven} and~\cite{bds-tuning} respectively. Intra-train IP feedback simulations with ground motion in the BDS are presented in~\cite{ryan}.

\subsection{Contributions of this Paper}
Beam-beam interactions in a linear collider can be strongly modified by correlations in the colliding beams. A well known example of this is the banana effect~\cite{banana3}. This motivates the need for start-to-end simulations that integrate each section of a linear collider. In this paper, we present for the first time integrated simulations of the RTML, ML and BDS of CLIC at 380\,GeV. We report a comprehensive study of the impact of static and dynamic imperfections on this collider. The following improvements have been made to the studies of static imperfections referenced in the previous section:
\begin{itemize}
\item We simulate a more complete list of static imperfections. Specifically in the ML, we now include magnet strength errors and beam position monitor (BPM) rolls.
\item We use the latest lattice, which is that submitted to the 2018 European Strategy for Particle Physics Update. Changes include a re-optimisation of the RF systems in the RTML~\cite{clic-pip} and an updated ML lattice~\cite{clic-pip}.
\item Simulation of the BDS collimation section. Most tuning simulations of the BDS focused on the final-focus system. In this work, we also simulate static imperfections in the collimation section.
\item Updated tuning procedures, in particular for the RTML and BDS. These are discussed further in Sec.\,\ref{s:tuning-procedure}.
\end{itemize}
We also study the impact of dynamic imperfections on this collider. For the first time, we perform integrated simulations including ground motion and stray magnetic fields from natural sources. Additional details of these simulations can be found in~\cite{thesis}. Tolerances for dynamic errors such as beam jitter, RF phase stability and magnet strength ripples are also calculated.

\subsection{Outline}
Details of the simulations performed in this work are given in Sec.\,\ref{s:integrated-simulations}. A perfect beamline is simulated in Sec.\,\ref{s:perfect-beamline}. Following this, the impact of different imperfections is studied. Sec.\,\ref{s:static-imperfections} looks at integrated simulations with static imperfections and the effectiveness of tuning procedures in recovering luminosity. Sec.\,\ref{s:dynamic-imperfections} looks at dynamic imperfections. Here, tolerances for dynamic errors are presented along with integrated simulations of ground motion and stray magnetic fields. Future work is discussed in Sec.\,\ref{s:next-steps} and the luminosity performance of CLIC at 380\,GeV is summarised in Sec.\,\ref{s:summary}.

\section{Integrated Simulations} \label{s:integrated-simulations}
In an integrated simulation, the beam is tracked from the exit of the DR to the IP. The simulation codes used in this work and the beam extracted from the DR are described below.

\subsection{Simulation Codes}
The particle tracking code PLACET~\cite{placet} was used to transport each beam from the DR to the IP. The tracking simulations include the emission of synchrotron radiation and short-range wakefields in the accelerating cavities.

A single bunch-crossing luminosity was calculated with a full simulation of the collision with the beam-beam effects code GUINEA-PIG~\cite{guinea}. This was multiplied by the repetition frequency and number of bunches per train to calculate the total luminosity.

The luminosity calculated with GUINEA-PIG is sensitive to the particle distribution of the colliding bunches. A small number of macro-particles leads to a high variance in the calculated luminosity. It was found that using 100,000 macro-particles leads to a standard deviation of less than 3\% of the mean value. In this paper, the luminosity is calculated with several hundred different beam distributions at the IP. The mean luminosity will be given. Each IP distribution was calculated with a tracking simulation in PLACET by sampling a new beam from the DR.

\begin{table}[!htb]
\centering
\begin{tabular}{l c c c}
\toprule
\textbf{Parameter} & \textbf{Symbol} & \textbf{Value} & \textbf{Unit} \\
\midrule
Horizontal/vertical emittance & $\epsilon_x/\epsilon_y$ & 700/5 & nm \\
Horizontal/vertical beam size & $\sigma_x/\sigma_y$ & 50/2.1 & $\mu$m \\
Horizontal/vertical beam divergence & $\sigma_{x'}/\sigma_{y'}$ & 2.5/0.4 & $\mu$rad \\
Bunch length & $\sigma_z$ & 1800 & $\mu$m \\
Energy & $E$ & 2.86 & GeV \\
Energy spread & $\sigma_E$ & 0.11 & \% \\
\bottomrule
\end{tabular}
\caption{\small Parameters of the beam extracted from the DR.}
\label{t:dr-beam}
\end{table}

\subsection{DR Beam}
In simulations, a Gaussian beam with 100,000 macro-particles is extracted from the DR. The simulated beam parameters at the exit of the DR are summarised in Table~\ref{t:dr-beam}.

\section{Perfect Beamline Performance} \label{s:perfect-beamline}
The maximum luminosity obtainable with this design of CLIC can be calculated by simulating a perfect collider. The luminosity achieved with a perfect collider is
\begin{equation}
\mathcal{L} = 4.3 \times 10^{34}~\text{cm}^{-2}\text{s}^{-1}.
\label{e:perfect-lumi}
\end{equation}
This is almost three times the nominal luminosity target (Eq.\,\eqref{e:nominal-luminosity-target}).

The beam parameters at the end of each section are shown in Table~\ref{t:perfect-beam-parameters}. There is an emittance growth of approximately 85\,nm in the horizontal direction and 0.8\,nm in the vertical direction that occurs in the RTML. This is from coherent and incoherent synchrotron radiation in the bends~\cite{rtml-csr}. A very small amount of emittance growth occurs in the ML due to imperfect matching to the RTML. The emittance growth in the BDS is due to correlations in the beam, which are described below.

\begin{table}[!htb]
\centering
\begin{tabular}{l c c c c c c c}
\toprule
\textbf{Section} & $\boldsymbol{\epsilon_x}$ \textbf{[nm]} &  $\boldsymbol{\epsilon_y}$ \textbf{[nm]} & $\boldsymbol{\sigma_x}$ \textbf{[$\boldsymbol{\mu}$m]} & $\boldsymbol{\sigma_y}$ \textbf{[$\boldsymbol{\mu}$m]} & $\boldsymbol{\sigma_z}$ \textbf{[$\boldsymbol{\mu}$m]} & $\boldsymbol{E}$ \textbf{[GeV]} & $\boldsymbol{\sigma_E}$ \textbf{[\%]} \\
\midrule
RTML & 785 & 5.82 & 18.9 & 0.63 & 70 & 9.00 & 1.0  \\
ML & 791 & 5.85 & 8.05 & 0.29 & 70 & 190 & 0.35 \\
BDS & 2,220 & 6.36 & 0.13 & 0.0013 & 70 & 190 & 0.35 \\
\bottomrule
\end{tabular}
\caption{\small Simulated beam parameters at the end of each section for perfect beamline.}
\label{t:perfect-beam-parameters}
\end{table}

\begin{figure}[!htb]
\centering
\includegraphics[width=0.3\linewidth]{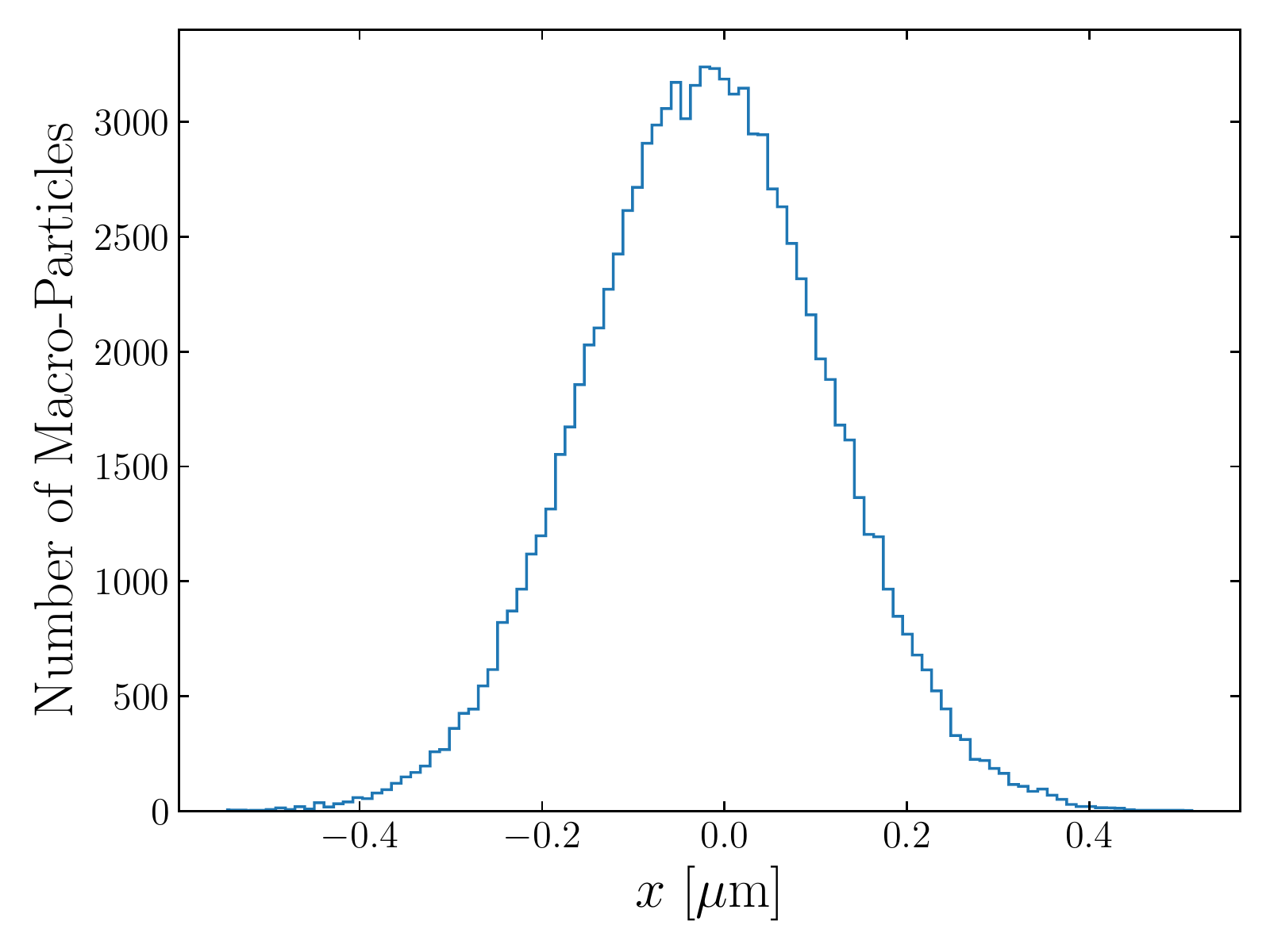}
\includegraphics[width=0.3\linewidth]{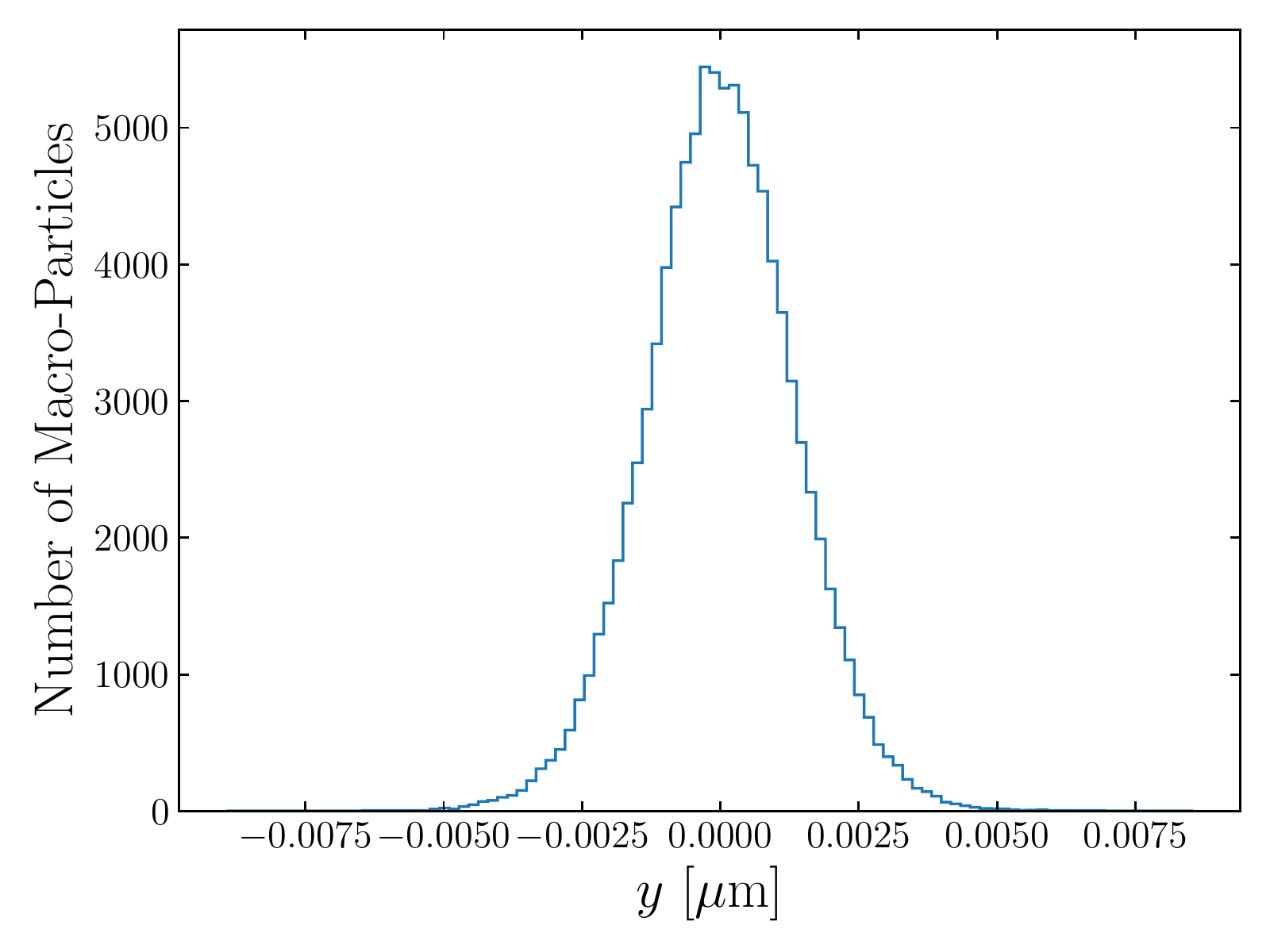}
\includegraphics[width=0.3\linewidth]{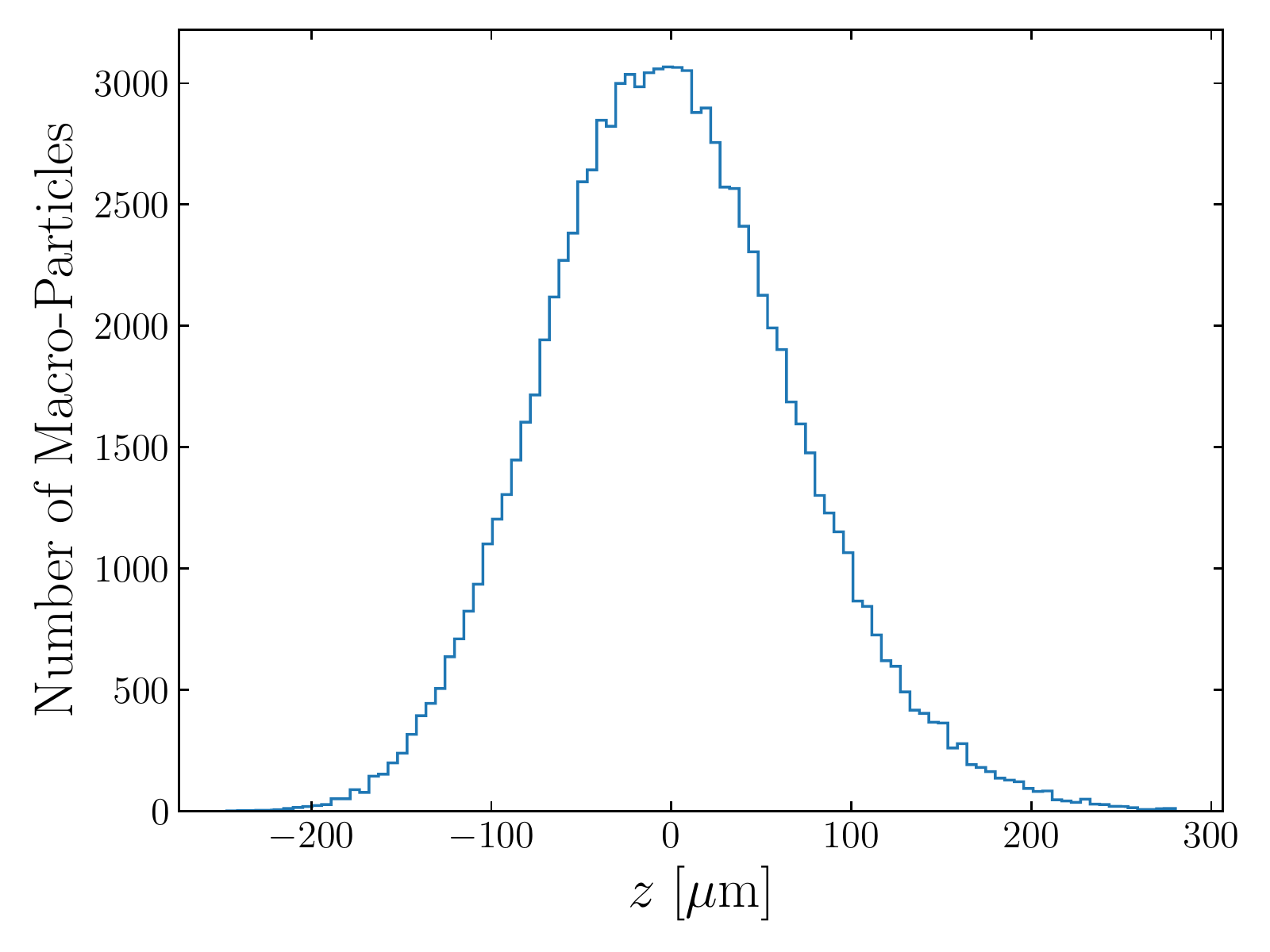}
\includegraphics[width=0.3\linewidth]{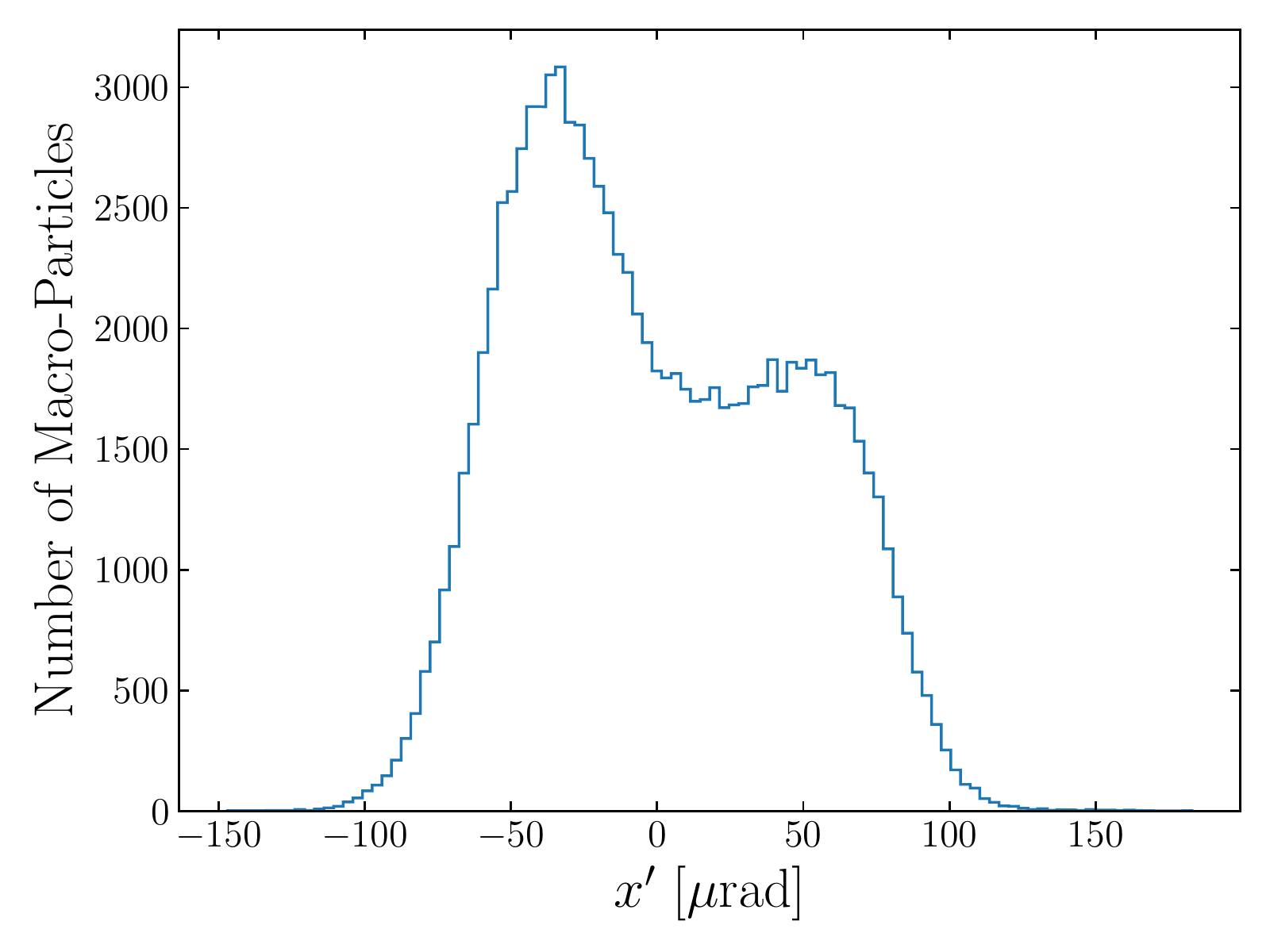}
\includegraphics[width=0.3\linewidth]{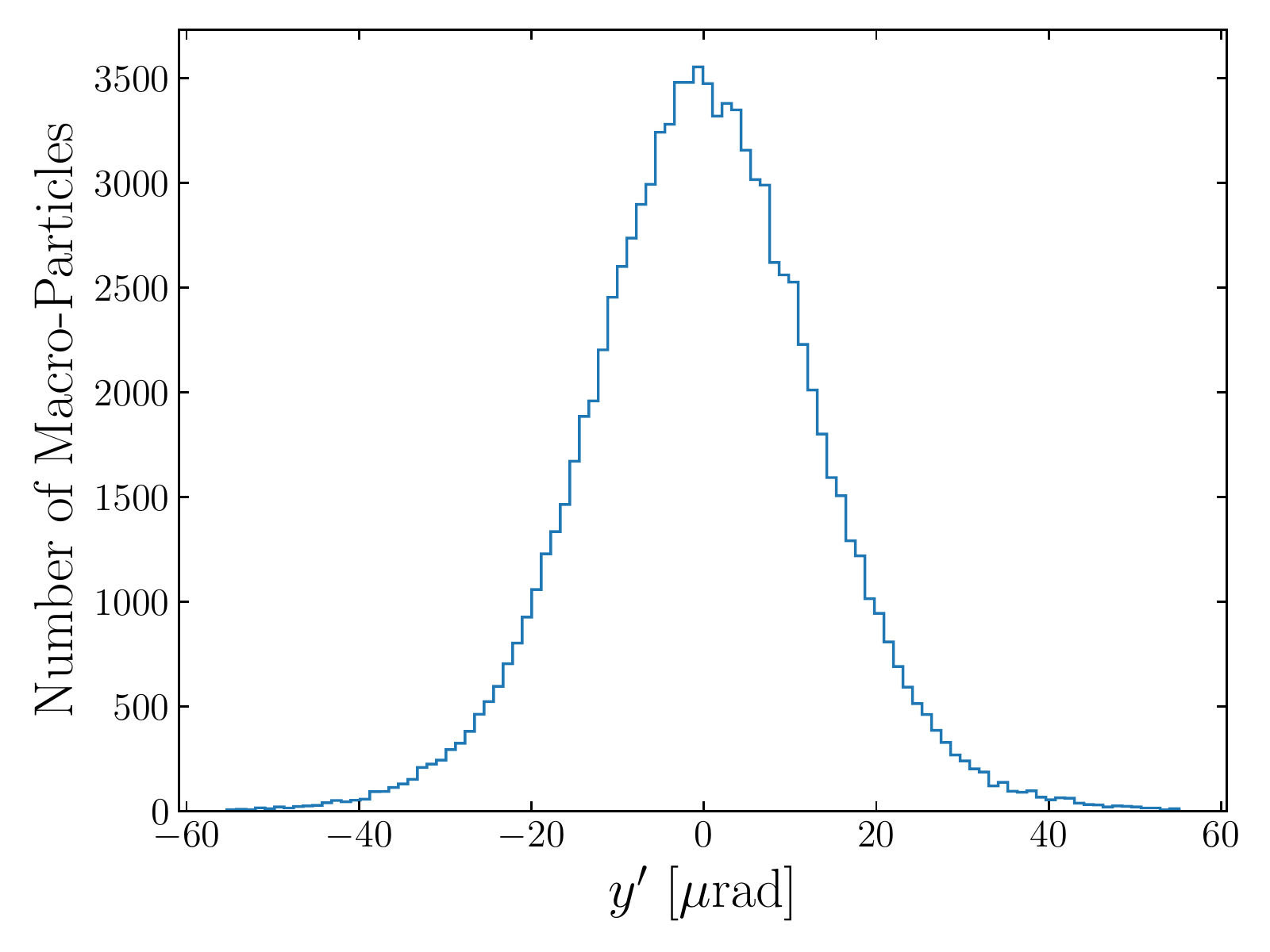}
\includegraphics[width=0.3\linewidth]{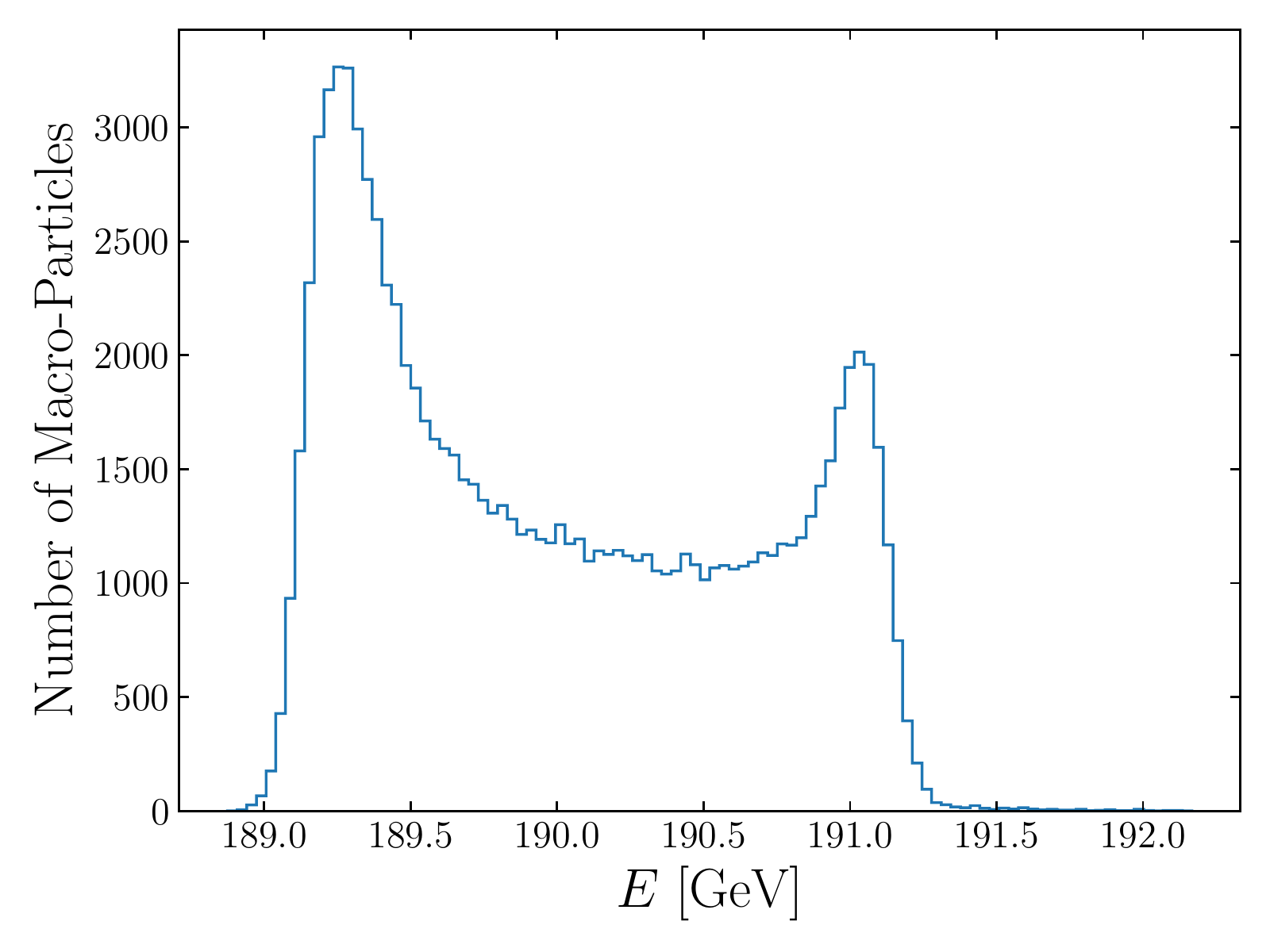}
\caption{\small Histogram of the horizontal position $x$ (top left), vertical position $y$ (top centre), longitudinal position $z$ (top right), horizontal angle $x'$ (bottom left), vertical angle $y'$ (bottom centre) and energy $E$ (bottom right) of a beam tracked through a perfect beamline.}
\label{f:ip-beam}
\end{figure}

\begin{figure}[!htb]
\centering
\includegraphics[width=0.55\linewidth]{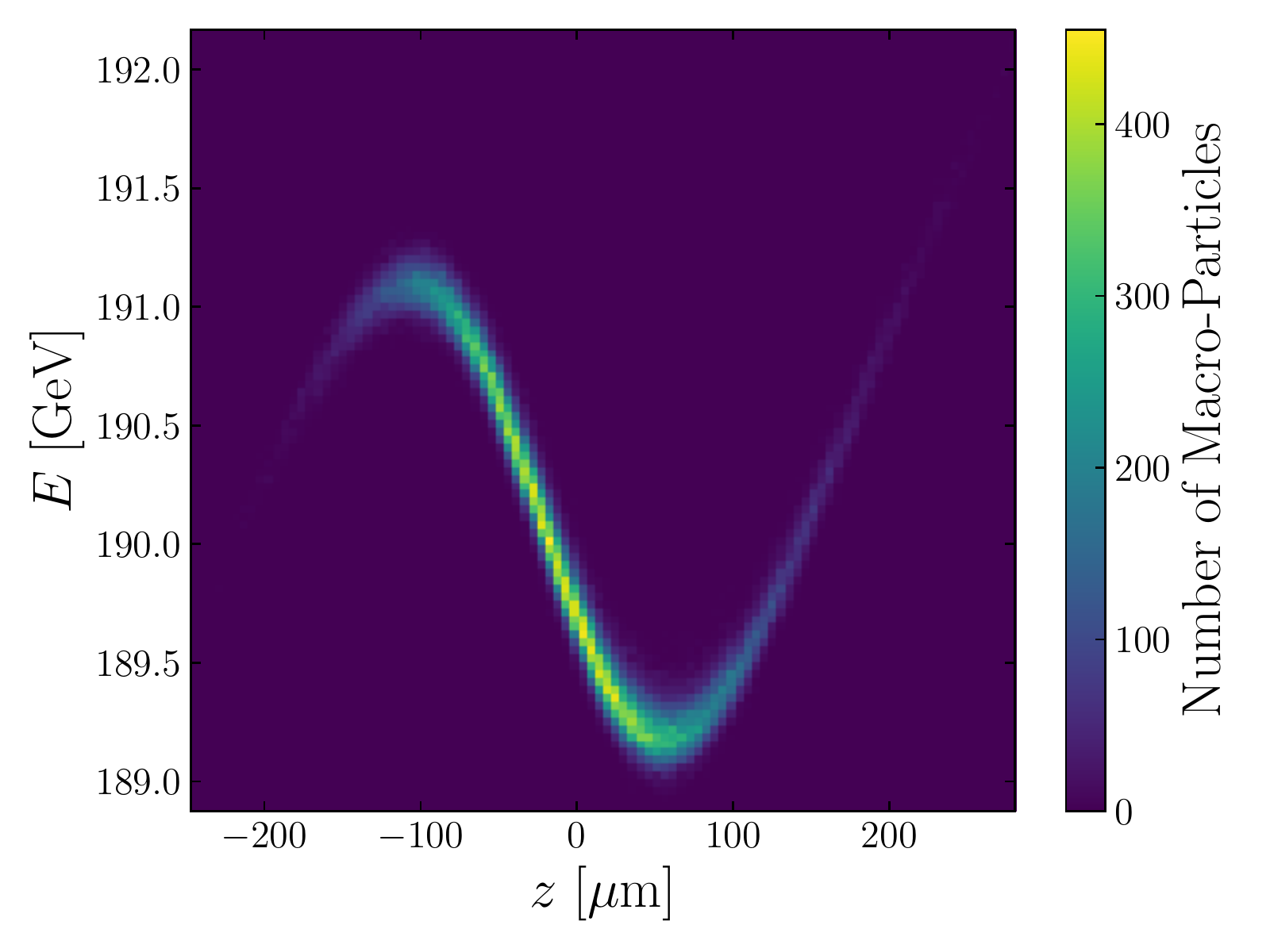}
\caption{\small Energy $E$ vs longitudinal position $z$ of the IP beam distribution. This beam was tracked through a perfect beamline.}
\label{f:ip-beam-zE}
\end{figure}

\begin{figure}[!htb]
\centering
\includegraphics[width=0.55\textwidth]{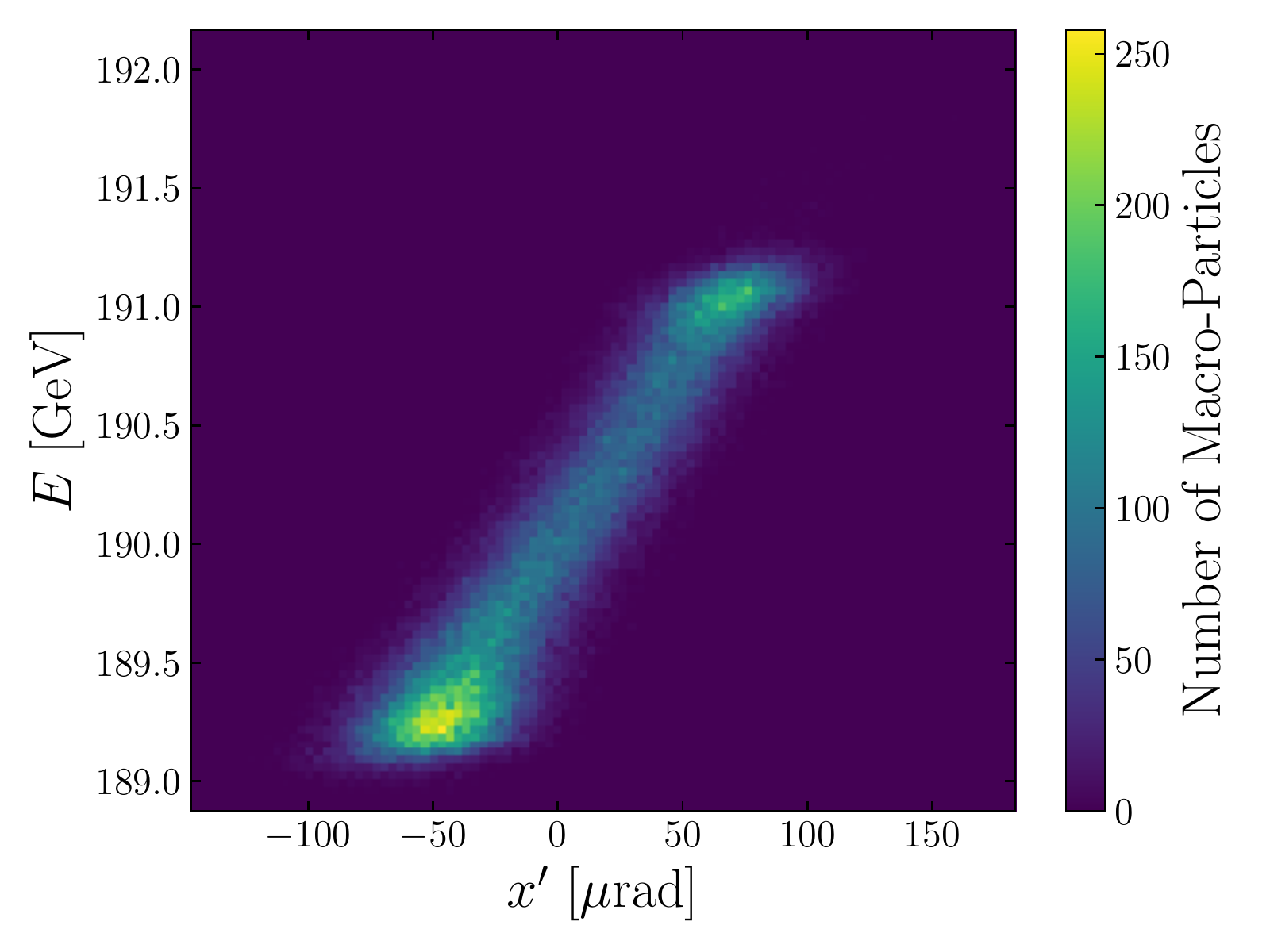} 
\caption{\small Energy $E$ vs horizontal angle $x'$ of the IP beam distribution. This beam was tracked through a perfect beamline.}
\label{f:ip-beam-energy-angle}
\end{figure}

Fig.\,\ref{f:ip-beam} shows the IP beam distribution generated by tracking a beam through a perfect beamline. There are two correlations in the IP beam distribution: between the $z$-$E$ and $x'$-$E$ coordinates. All other coordinates are uncorrelated. The $z$-$E$ and $x'$-$E$ correlations are shown in Figs.\,\ref{f:ip-beam-zE} and \ref{f:ip-beam-energy-angle} respectively. The $z$-$E$ correlation arises from short-range wakefields in the ML cavities and from off-crest acceleration, which is optimised to minimise the energy spread at the end of the ML without compromising beam stability.

In the BDS, sextupoles are placed in dispersive regions to correct chromaticity~\cite{local-chromaticity-correction}. This results in a correlation between the energy and horizontal angle. This can be seen in Figs.\,\ref{f:ip-beam} and \ref{f:ip-beam-energy-angle}. The correlation leads to a horizontal emittance growth in the BDS. As this emittance growth is from the angular distribution, it does not significantly impact the IP beam size and luminosity.

\section{Static Imperfections} \label{s:static-imperfections}
\begin{figure}[!htb]
\centering
\includegraphics[width=0.32\textwidth]{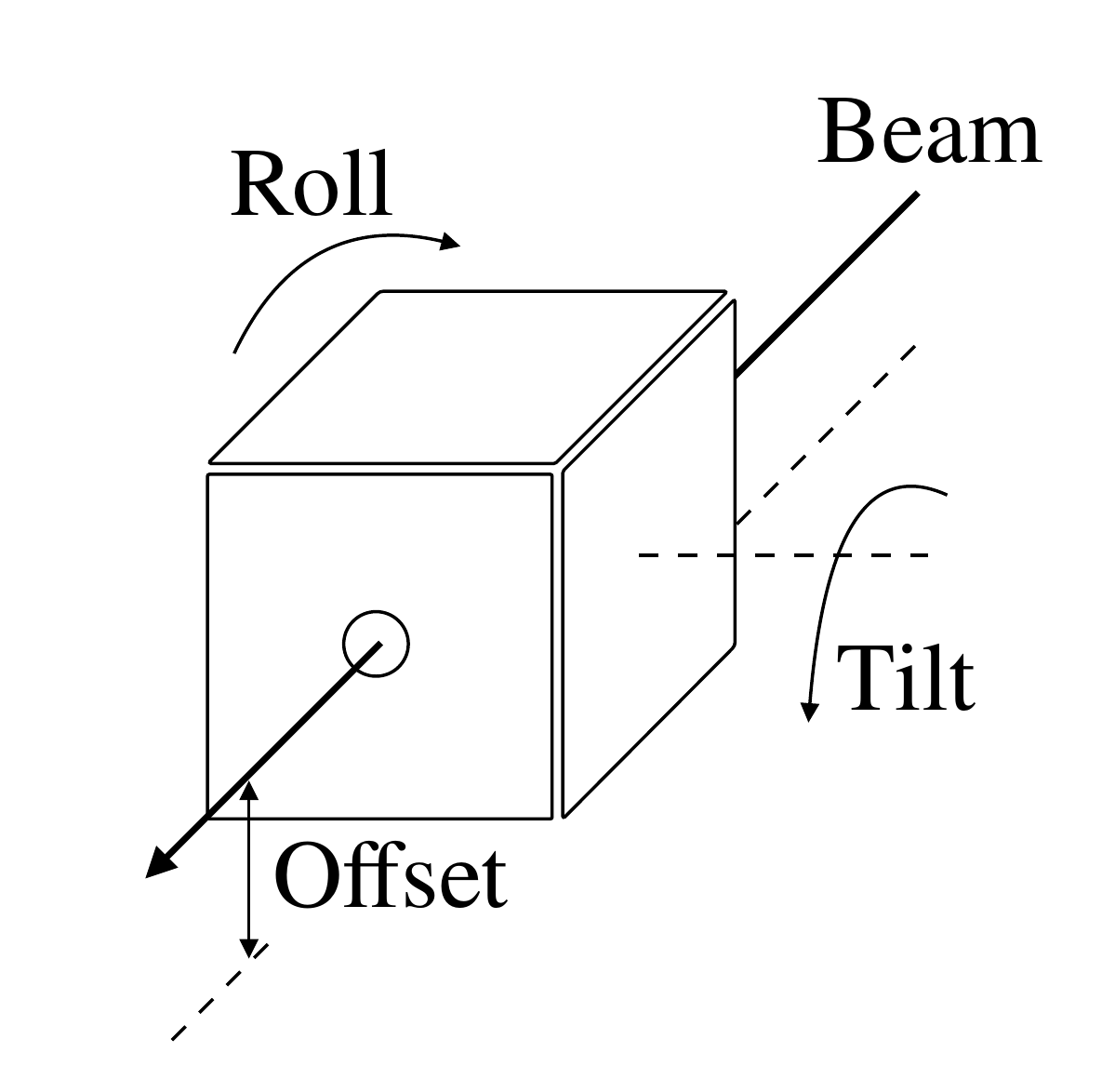} 
\caption{\small Illustration of different types of misalignments: roll, tilt and offset. In each case, the dashed line is the reference.}
\label{f:misalignments}
\end{figure}

Static imperfections include errors in the alignment of accelerator elements, which are illustrated in Fig.\,\ref{f:misalignments}, and static errors in the attributes of elements.

\subsection{Description}\label{s:static-error-description}
\subsubsection{Beam Position Monitors}
The most important static imperfection is the misalignment of BPMs. BPMs define the ideal trajectory of a beam. Therefore, if they are offset with respect to a straight-line trajectory, the beam will not follow a straight path. Furthermore, if the BPMs are rolled, a horizontal beam offset will appear partially as a vertical offset and vice versa, which complicates the centring of the beam in each plane.

Additionally, an important static imperfection is the noise of a BPM. Each BPM reading is corrupted by an error, which in simulations is assumed to be Gaussian. The standard deviation of this error is the BPM resolution. Corrections are applied to the beam based on BPM readings. Therefore, a good resolution is desired to minimise the introduction of noise from the BPM readings to the beam.

\subsubsection{Accelerating Cavities}
The misalignment of accelerating cavities is another important static imperfection. A cavity offset with respect to the beam excites wakefields, which lead to emittance growth. Novel wakefield monitors~\cite{wakefield-monitors} are used to measure the wakefield in CLIC cavities.

Additionally, tilts are an important alignment error for cavities. If a cavity is tilted, a component of the accelerating voltage is applied in a transverse direction with respect to the beam. This results in the beam being kicked.

\subsubsection{Magnets}
Important static imperfections for magnets are strength errors and misalignments with respect to the ideal beam. Offset quadrupole and sextupole magnets kick the beam and lead to emittance growth. Additionally, magnet rolls lead to an $xy$-coupling, which results in emittance growth.

\subsubsection{Girders}
CLIC will utilise the pre-alignment procedure described in~\cite{clic-cdr}. Elements are placed on girders, which are attached to movers equipped with sensors. A system of stretched wires is used as a reference to align elements to a root-mean-square (RMS) offset of 10\,$\mu$m over distances of 200\,m. Girders can be misaligned with respect to the reference line and articulation points.

\subsubsection{Summary}
All errors are assumed to have a Gaussian distribution. A summary of the errors simulated in each section of CLIC is given in Table~\ref{t:static-imperfections}. In PLACET, misalignments are simulated with respect to a perfect straight-line trajectory, which in reality corresponds to the system of wires used as the reference.

The imperfections listed in Table~\ref{t:static-imperfections} are based on previous tuning studies and have been defined in discussion with instrumentation, magnet and RF experts~\cite{clic-pip}. The RMS errors listed for the RTML have been achieved or exceeded in existing accelerator facilities~\cite{clic-pip}. The errors in the ML have been deemed achievable by experts~\cite{clic-pip}. The requirements in the ML and BDS are the same for the 380\,GeV and 3\,TeV stages of CLIC to avoid system upgrades in later stages~\cite{clic-pip, clic-cdr}.

\begin{table*}[!htb]
\centering
\begin{tabular}{l l c}
\toprule
& \textbf{Imperfection} & \textbf{Value} \\
\midrule
RTML & & \\
\midrule
& Magnet and BPM offset & 30 $\mu$m \\
& Magnet and BPM roll & 100 $\mu$rad \\
& BPM resolution & 1 $\mu$m \\
& CA and TA quadrupole strength errors & 0.01\% \\
& All other magnet strength errors & 0.1\% \\
\midrule
ML & & \\
\midrule
& Magnet and BPM offset & 14 $\mu$m \\
& Magnet and BPM roll & 100 $\mu$rad \\
& BPM resolution & 0.1 $\mu$m \\
& Magnet strength errors & 0.01\% \\
& Girder end point with respect to reference wire & 12 $\mu$m \\
& Girder end point with respect to articulation point & 5 $\mu$m \\
& Accelerating structure offset &14 $\mu$m \\
& Accelerating structure tilt & 141 $\mu$rad \\
& Wakefield monitor offset & 3.5 $\mu$m \\
\midrule
BDS & & \\
\midrule
& Magnet and BPM offset & 10 $\mu$m \\
& Magnet and BPM roll & 100 $\mu$rad \\
& BPM resolution & 20 nm \\
& Magnet strength errors & 0.01\% \\
\bottomrule
\end{tabular}
\caption{\small RMS values for static imperfections implemented in integrated simulations. CA is the central arc and TA is the turn around (see Fig.\,\ref{f:clic-380-gev}).}
\label{t:static-imperfections}
\end{table*}

\subsection{Tuning Procedure}\label{s:tuning-procedure}
Following pre-alignment, several well known beam-based alignment methods are used to tune the beamline. These are described below.

\subsubsection{One-to-One (121) Steering}
This is the first tuning step. The beam is electrically centred in each BPM using the nearest upstream corrector. In the RTML dipoles are used to apply the correction. In the ML and BDS, quadrupoles mounted on movers are displaced to apply the correction.

The corrector settings $\boldsymbol{\theta}$ are found by minimising the objective function~\cite{bds-tuning, rtml-tuning2}
\begin{equation}
\chi^2 = | \boldsymbol{\Delta u} - \boldsymbol{R}\,\boldsymbol{\theta} |^2 + \beta^2_0 | \boldsymbol{\theta} |^2,
\end{equation}
where $\boldsymbol{\Delta u} = \boldsymbol{u} - \boldsymbol{u}_0$, $\boldsymbol{u}$ is a vector containing the BPM readings of a beam tracked through an imperfect beamline, $\boldsymbol{u}_0$ is a vector containing the BPM readings of an ideal beam, $\boldsymbol{R}$ is the response matrix and $|.|$ denotes the magnitude. $\beta_0$ is a free parameter that is included to avoid large corrector strengths.

\subsubsection{Dispersion-Free Steering (DFS)}
Following 121 steering, DFS is performed. Here, the correctors are used to minimise the difference in the trajectory of two beams of differing energy. The corrector settings are found by minimising the objective function~\cite{bds-tuning, rtml-tuning2}
\begin{equation}
\chi^2 = | \boldsymbol{\Delta u} - \boldsymbol{R}\,\boldsymbol{\theta} |^2 + \omega^2 |\boldsymbol{\eta} - \boldsymbol{D} \, \boldsymbol{\theta} |^2 + \beta^2_1 | \boldsymbol{\theta} |^2,
\end{equation}
where $\boldsymbol{\eta} = \boldsymbol{u}_{\Delta E} - \boldsymbol{u_0}$, $\boldsymbol{u}_{\Delta E}$ is a vector containing the BPM readings using an off-energy beam, $\boldsymbol{u}_0$ is the same as in the previous equation and $\boldsymbol{D}$ is the dispersion response matrix. $\beta_1$ is another free parameter to avoid large corrector settings and $\omega$ is a weight factor for the dispersion term, which can be calculated as~\cite{bds-tuning, rtml-tuning2}
\begin{equation}
\omega^2 = \frac{\sigma^2_\text{mis} + \sigma^2_\text{res}}{2\sigma^2_\text{res}},
\label{e:dispersion-weight}
\end{equation}
where $\sigma_\text{mis}$ is the RMS BPM offset and $\sigma_\text{res}$ is the BPM resolution. Usually, the optimum value for $\omega$ is slightly different to Eq.\,\eqref{e:dispersion-weight} due to non-linear effects, such as wakefields and synchrotron radiation. A scan is performed to find the optimum value for $\omega$. Values of $\omega$, $\beta_0$ and $\beta_1$ from~\cite{ml-tuning, bds-tuning, rtml-tuning2} were used in this work.

\subsubsection{Section-Specific Tuning}
Following 121 steering and DFS, specific tuning procedures are performed that depend on the section. 

\textbf{RF realignment} is performed in the ML. This involves offsetting a cavity to minimise the reading from a wakefield monitor~\cite{ml-tuning}.

\textbf{Sextupole tuning} for chromaticity correction is performed in the RTML and BDS. In the RTML, a simplex algorithm~\cite{simplex} is used to minimise the emittance at the end of the section by moving the last five sextupoles in the central arc and the last five sextupoles in the turn-around loop.

Previous tuning studies for the RTML simultaneously minimised the horizontal and vertical emittance at the end of the section (see~\cite{rtml-tuning2}). This procedure would often find a solution that minimised the vertical emittance only. Here, the horizontal and vertical emittances were minimised separately. This produced beams with a lower horizontal and vertical emittance compared to the previous procedure.

The BDS collimation section uses a simplex algorithm that displaces sextupoles in order to minimise the emittance at the start of the final-focus system.

For the final-focus system, the tuning signal used is the luminosity instead of the emittance. Two beams are required to calculate a luminosity. Ideally, each beam would be tracked through its own beamline. However, at the time of these studies the latest tuning procedures for the final-focus system simulated a single beam and mirrored it at the IP to calculate a luminosity. This method of calculating the luminosity was only used for tuning. To estimate the luminosity of the collider two beams were tracked through different tuned beamlines, this is discussed further in Sec.\,\ref{s:static-lumi-opt}.

A combination of a random walk of sextupole offsets and sextupole knobs is used to tune the final-focus system after 121 steering and DFS. These procedures are described in~\cite{bds-tuning}. When tuning the final-focus system a small number of cases get trapped in a local optimum, which prevents them from reaching the maximum possible luminosity. A new step is introduced for these beamlines: a random walk of quadrupole and sextupole offsets~\cite{bds-tuning}. This puts the beamlines into a new configuration, which can make the sextupole tuning knobs more effective. Reapplying the tuning knobs further increases the luminosity of these beamlines.

\subsubsection{Luminosity Optimisation} \label{s:static-lumi-opt}
The beam-based tuning procedures described above were applied to 100 beamlines containing static imperfections. We use each of these beamlines to track the electron beam. We then randomly select a beamline from the remaining 99 beamlines to track the positron beam. Therefore, we have 100 unique beamline pairs, which we will refer to as colliders.

Each beamline in a collider has been tuned independently. Therefore, the achieved luminosity is not necessarily the optimum. Furthermore, for colliders that have a high disruption, correlations in the beam can influence the luminosity and result in the maximum luminosity occurring with a beam-beam offset. To optimise the luminosity, we perform a vertical beam-beam offset scan, a vertical crossing angle scan and waist scan. Optimising the horizontal beam-beam offset and crossing angle had a small effect (less than 1\% luminosity gain) so was not included in the tuning procedure. Because each beamline was tuned independently, the luminosity that is achieved is a conservative estimate, which may be improved by performing two-beam tuning.

\subsection{Luminosity}
\begin{figure}[!htb]
\centering
\includegraphics[width=0.55\textwidth]{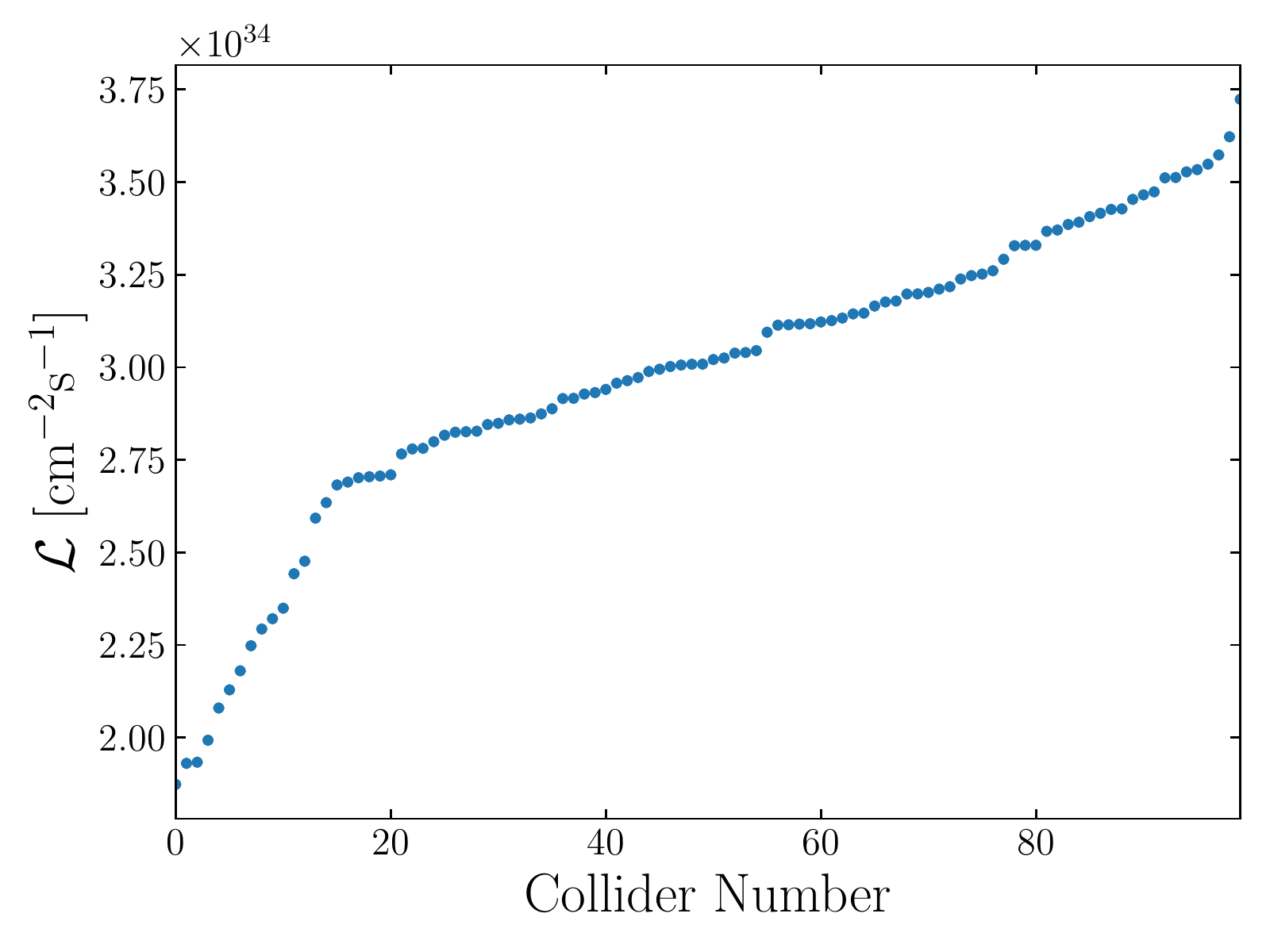}
\caption{\small Luminosity $\mathcal{L}$ vs collider number for 100 tuned colliders with static imperfections. Colliders are ordered in ascending luminosity.}
\label{f:tuned-luminosity}
\end{figure}

Fig.\,\ref{f:tuned-luminosity} shows the luminosity of 100 colliders after the full tuning procedure. The mean luminosity and its standard deviation is
\begin{equation}
\mathcal{L} = (3.0 \pm 0.4) \times 10^{34}~\text{cm}^{-2}\text{s}^{-1}.
\label{e:mean-lumi-static}
\end{equation}
90\% of colliders achieve a luminosity greater than $2.35\times10^{34}\,\text{cm}^{-2}\text{s}^{-1}$, which expressed as a percentage of the nominal luminosity target is 157\%. This means there is a significant surplus of 57\%, providing a margin for the impact of dynamic imperfections.

\section{Dynamic Imperfections} \label{s:dynamic-imperfections}
This section reviews short-term dynamic imperfections. These are processes that impact the beam on a train-to-train basis, which are difficult to correct because of their fast temporal variation. The most important dynamic imperfections for CLIC are beam jitter, RF phase errors, magnetic field ripples, ground motion and stray magnetic fields.

\subsection{Tolerances}
For beam jitter, phase errors and magnetic field ripples, tolerances to limit luminosity loss are calculated. These are presented below.

\begin{figure}[!htb]
\centering
\includegraphics[width=0.55\textwidth]{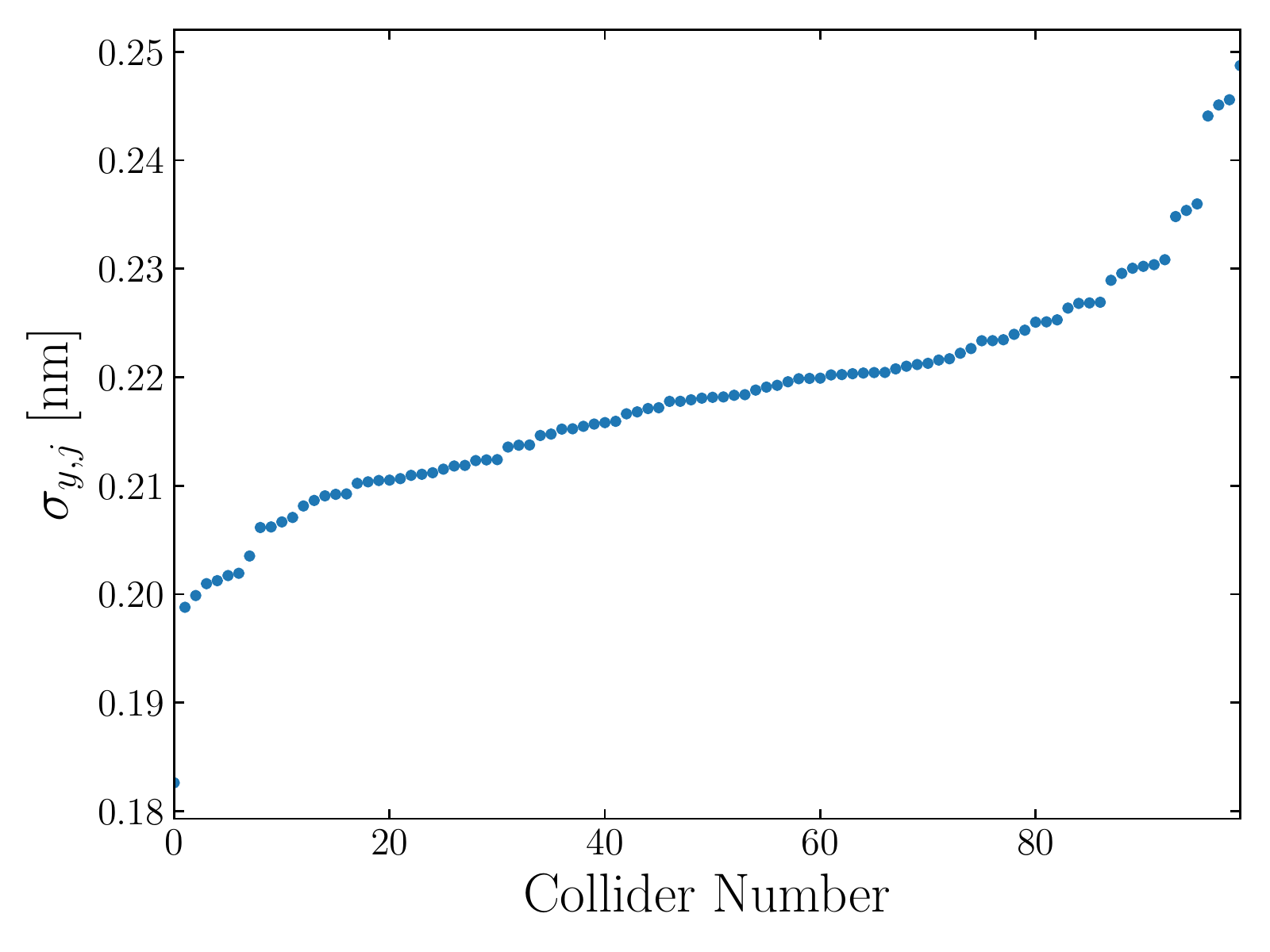}
\caption{\small Vertical position jitter at the IP $\sigma_{y,j}$ that corresponds to a luminosity loss of $3\times10^{32}\,\text{cm}^{-2}\text{s}^{-1}$ for 100 tuned colliders with static imperfections. Colliders are ordered in ascending position jitter.}
\label{f:beam-jitter}
\end{figure}

\subsubsection{Beam-Beam Offset and IP Beam Jitter}
The luminosity with a vertical beam-beam offset is influenced by beam-beam interactions, which themselves depend on many factors, such as correlations in the beam and the IP emittance. Each collider has its own sensitivity to luminosity loss due to a beam-beam offset. A vertical position jitter $\sigma_{y,j}$ was simulated at the IP for each beam. Fig.\,\ref{f:beam-jitter} shows the vertical position jitter that corresponds to a luminosity loss of $3\times10^{32}\,\text{cm}^{-2}\text{s}^{-1}$ (2\% of the nominal luminosity target) for 100 tuned colliders. The mean vertical position jitter tolerance and its standard deviation is
\begin{equation}
\sigma_{y,j} = (0.22 \pm 0.01)~\text{nm}.
\end{equation}
90\% of colliders have a tolerance greater than 0.20\,nm. The IP beam jitter is largely determined by the stability of the final doublet, which measurements have shown can be stabilised to an RMS jitter of less than 0.2\,nm~\cite{clic-cdr}. It is also possible to relax this tolerance by including an intra-train IP feedback, such as the FONT system~\cite{font, font2}.

\subsubsection{RF Phase Errors}
This section examines the impact of coherent RF phase errors. For incoherent RF phase errors, there is an averaging effect that generally leads to much larger tolerances, which means they are less important~\cite{dynamic-effect-clic-ml}.

\begin{figure}[!htb]
\centering
\includegraphics[width=0.55\textwidth]{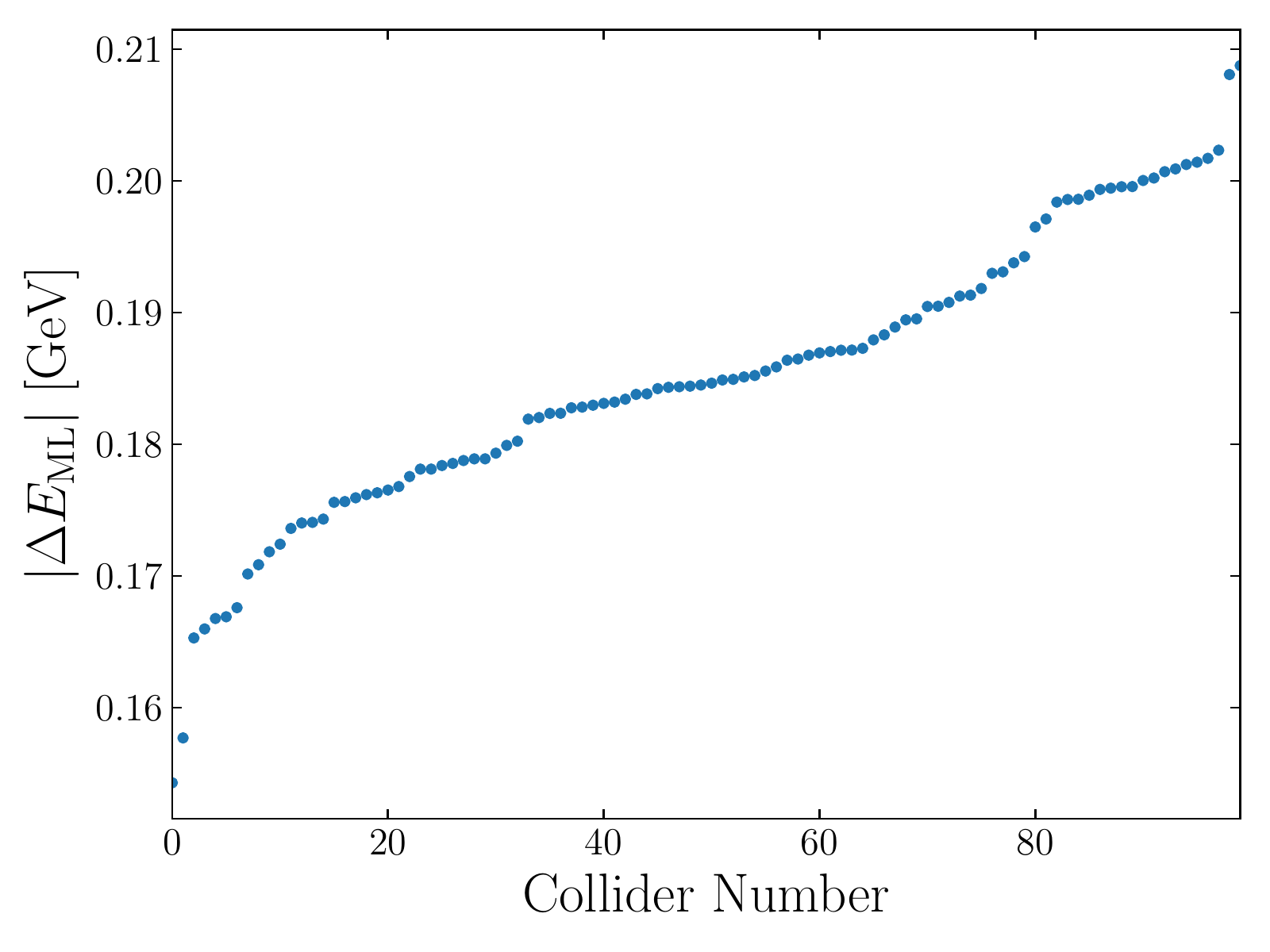}
\caption{\small ML beam energy error $|\Delta E_\text{ML}|$ that corresponds to a luminosity loss of $1.5\times10^{32}\,\text{cm}^{-2}\text{s}^{-1}$ vs collider number for 100 tuned colliders with static imperfections. Colliders are ordered in ascending energy error.}
\label{f:phase-errors}
\end{figure}

Coherent RF phase errors in the ML cavities are equivalent to a global error in the accelerating gradient. This leads to an off-energy beam at the end of the ML. Due to chromaticity, an off-energy beam in the BDS has a larger beam size and yields a lower luminosity. Fig.\,\ref{f:phase-errors} shows the energy error at the end of the ML that corresponds to a luminosity loss of $1.5\times10^{32}\,\text{cm}^{-2}\text{s}^{-1}$ (1\% of the nominal luminosity target) for 100 tuned colliders. The energy of the beam was varied by changing the effective gradient of the ML cavities. The RF phase errors were only applied to one of the beamlines. The mean tolerance and its standard deviation is
\begin{equation}
|\Delta E_\text{ML}| = (0.19 \pm 0.01)~\text{GeV}.
\end{equation}
This corresponds to an RF phase error of $\pm 0.29^\circ$ in the ML cavities, which have an RF frequency of 12\,GHz. 90\% of colliders have a tolerance greater than $\pm 0.17$\,GeV, which is equivalent to an RF phase error of $\pm 0.26^\circ$ in the ML cavities. In CLIC, the RF phase stability is determined by the arrival time of the drive beam, which has a demonstrated stability of $(0.20 \pm 0.01)^\circ$~\cite{jack}.

\subsubsection{Quadrupole Ripples}
Magnetic field ripples arise from power supply ripples. In simulations, a relative RMS error $\sigma_B /B$ was applied to the strength of every quadrupole. A tolerance was chosen for each section as the relative RMS error that results in a luminosity loss of less than $1.5\times10^{32}\,\text{cm}^{-2}\text{s}^{-1}$ (1\% of the nominal luminosity target). These tolerances are presented in Table~\ref{t:magnet-stability}. The BDS, particularly the final doublet, has the tightest requirements. The tightest tolerances for CLIC are similar to those found in the Large Hadron Collider~\cite{lhc-design-report}.

\begin{table}[!htb]
\centering
\begin{tabular}{l c}
\toprule
\textbf{Section} & \textbf{$\boldsymbol{\sigma_B / B}$} \\
\midrule
RTML & $10^{-4}$ \\
ML & $10^{-4}$ \\
BDS (Excluding FD) & $10^{-5}$ \\
FD & $10^{-6}$ \\
\bottomrule
\end{tabular}
\caption{Quadrupole ripple tolerances $\sigma_B/B$ for specific sections. FD is the final doublet.}
\label{t:magnet-stability}
\end{table}

\subsection{Ground Motion}
This section describes models used to simulate ground motion and the mitigation systems used in CLIC to limit luminosity loss.

\subsubsection{Models}
Ground motion is modelled as a set of travelling waves with differing wavelength and frequency. The amplitude of these waves is determined by a 2D power spectral density (PSD)~\cite{seryi}. There are several models which specify a 2D PSD~\cite{clic-cdr, seryi}. Ground motion model D\footnote{Also known as model B10.} is studied in this work. Model D represents a higher level of ground motion than CLIC is expected to experience. It is based on measurements at SLAC~\cite{slac-measurements}, Fermilab~\cite{fermilab-measurements} and in the CMS detector cavern~\cite{cms-measurements}.

\begin{figure}[!htb]
\centering
\includegraphics[width=0.55\textwidth]{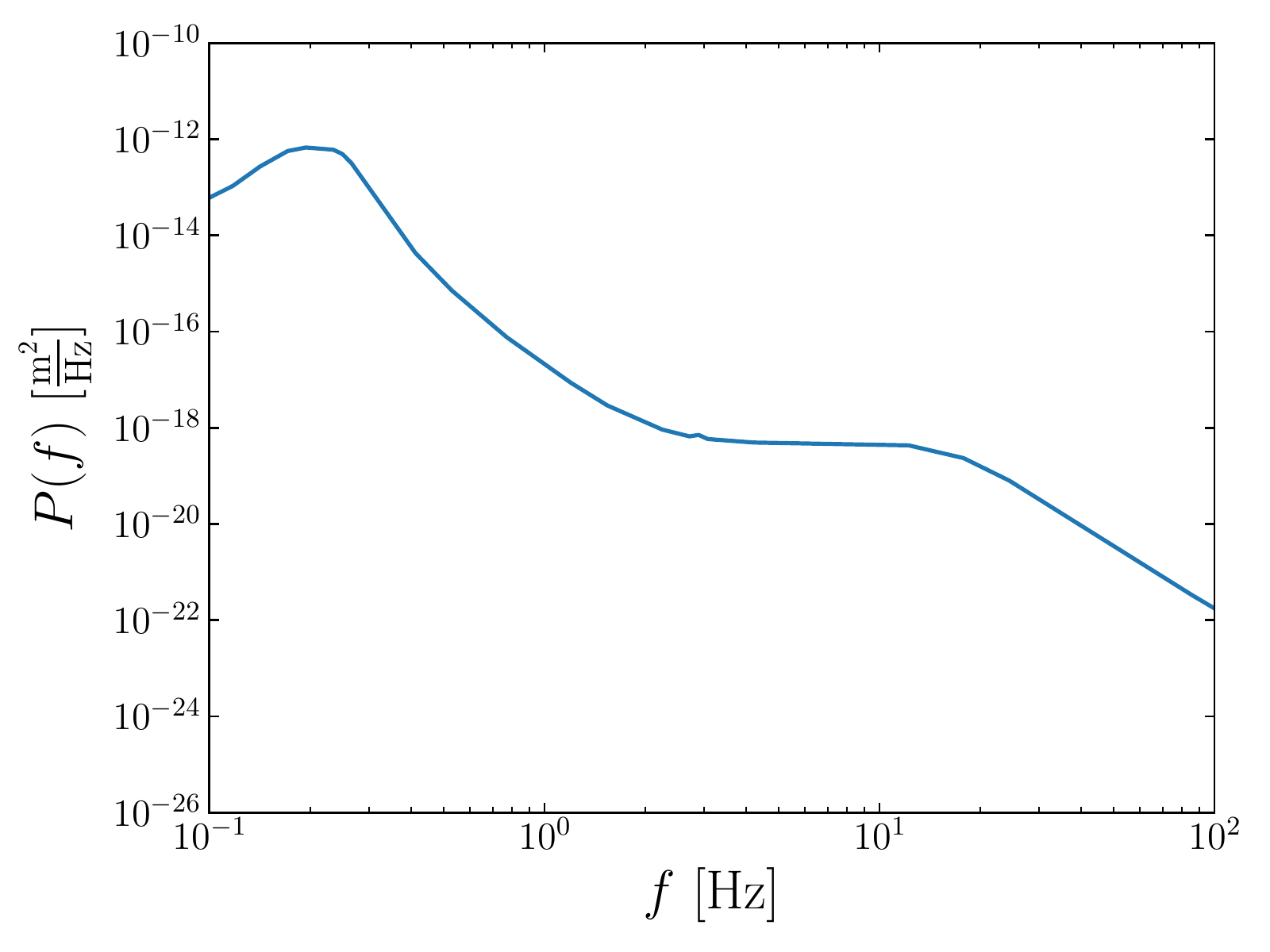} 
\caption{\small PSD $P(f)$ vs frequency $f$ of ground motion model D.}
\label{f:gm-models}
\end{figure}

\begin{figure}[!htb]
\centering
\includegraphics[width=0.55\textwidth]{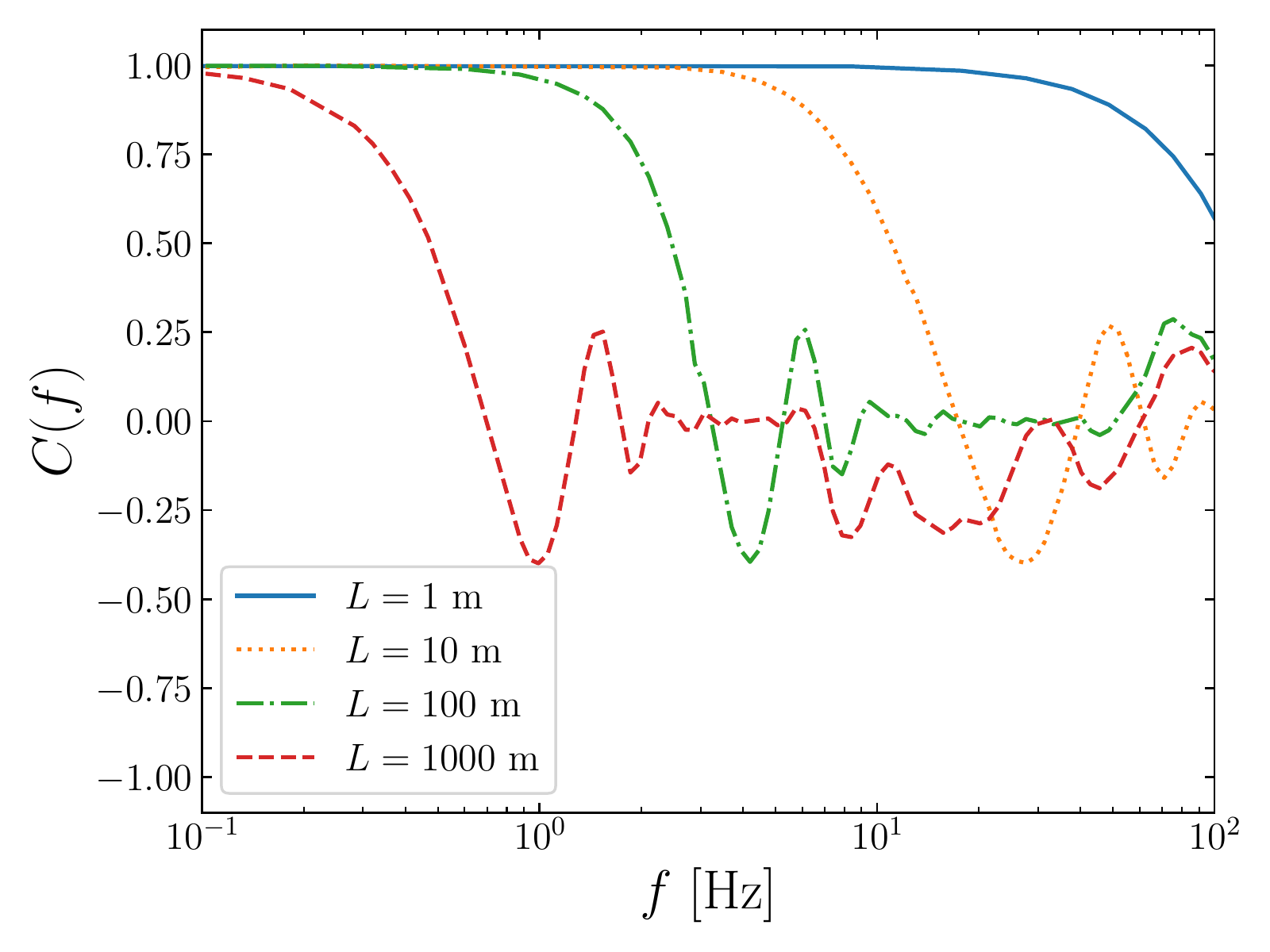} 
\caption{\small Correlation $C(f)$ vs frequency $f$ of ground motion model D for different separations $L$.}
\label{f:gm-cor}
\end{figure}

The 1D PSD of model D is shown in Fig.\,\ref{f:gm-models}. There are two broadband peaks in this PSD. One at 0.14\,Hz, which arises from ocean waves, and another at approximately 20\,Hz, which arises from technical equipment in the accelerator tunnel. The correlation of ground motion at different locations is shown in Fig.\,\ref{f:gm-cor}. Low frequencies have a high correlation across large distances, whereas high frequencies are only correlated over short distances.

\subsubsection{Mitigation Systems}
There are two systems that are essential to mitigate the impact of ground motion: a beam-based feedback system and a quadrupole stabilisation system. These are described below.

\textbf{The beam-based feedback system} aims to correct the beam offset. As described in~\cite{seryi}, the average impact of a beam-based feedback system can be estimated by applying a transfer function $T(f)$ to the ground motion PSD to give an effective PSD,
\begin{equation}
P_\text{eff}(f) = |T(f)|^2 P(f),
\end{equation}
which is then used in simulations to generate the ground displacement. This approach assumes that a perfect correction is applied by the feedback system and that the transfer function depends only on the frequency, i.e. the feedback system has the same effect across the entire accelerator. This is a simplification, however if the feedback control is designed well, this is a good approximation~\cite{jurgen, clic-feedback1, clic-feedback2}.

\begin{figure}[!htb]
\centering
\includegraphics[width=0.55\textwidth]{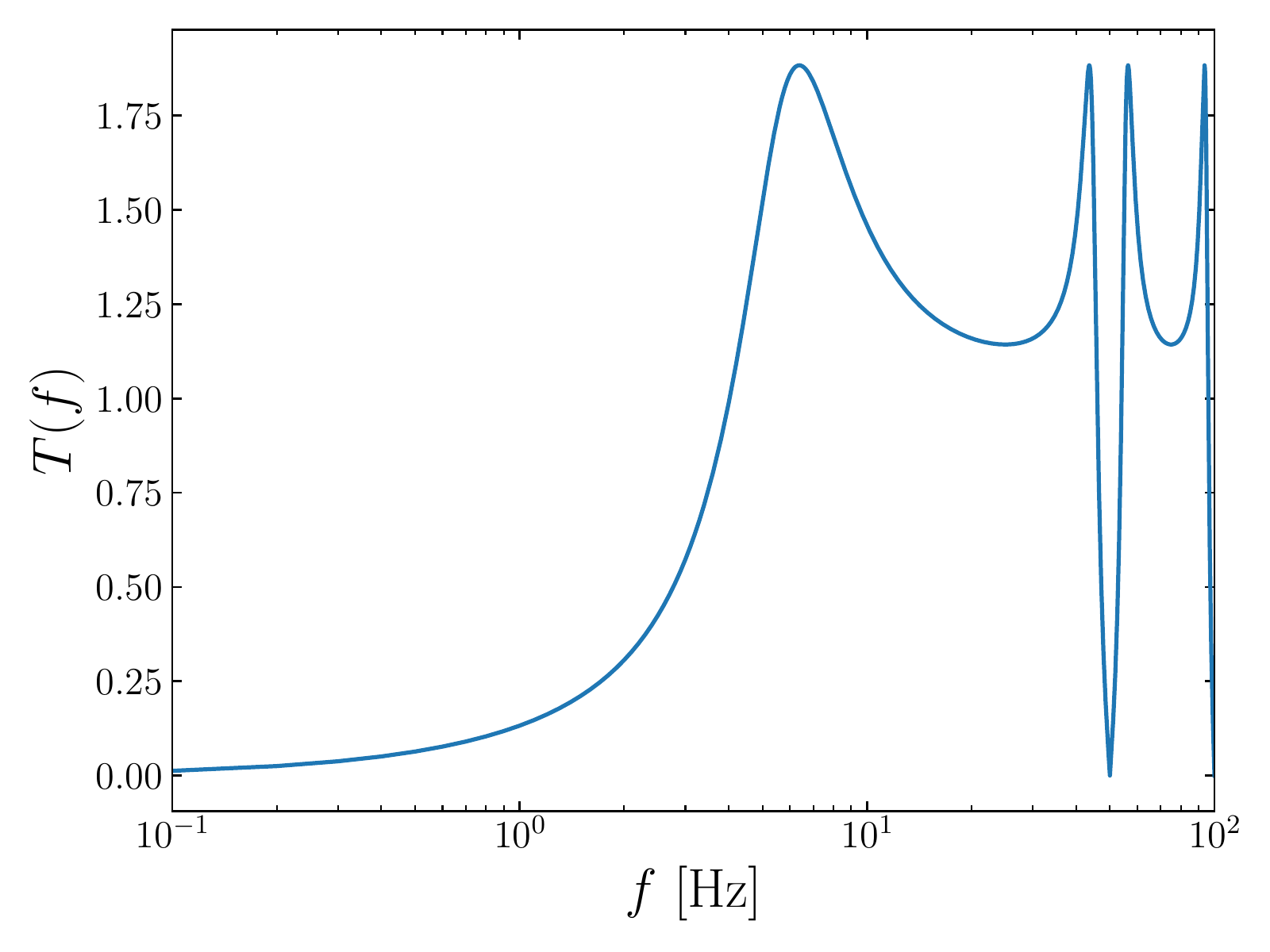} 
\caption{\small Transfer function $T(f)$ vs frequency $f$ for the beam-based feedback system used in CLIC.}
\label{f:feedback-tf}
\end{figure}

The transfer function of the beam-based feedback system used in CLIC is shown in Fig.\,\ref{f:feedback-tf}. The feedback system is effective for mitigating low frequencies, below 1\,Hz. It amplifies frequencies in the range 4-25\,Hz. The repetition frequency of the CLIC beam is 50\,Hz. Therefore, dynamic imperfections at harmonics of 50\,Hz appear static to the beam and the transfer function for the beam-based feedback system is zero.

\begin{figure}[!htb]
\centering
\includegraphics[width=0.55\textwidth]{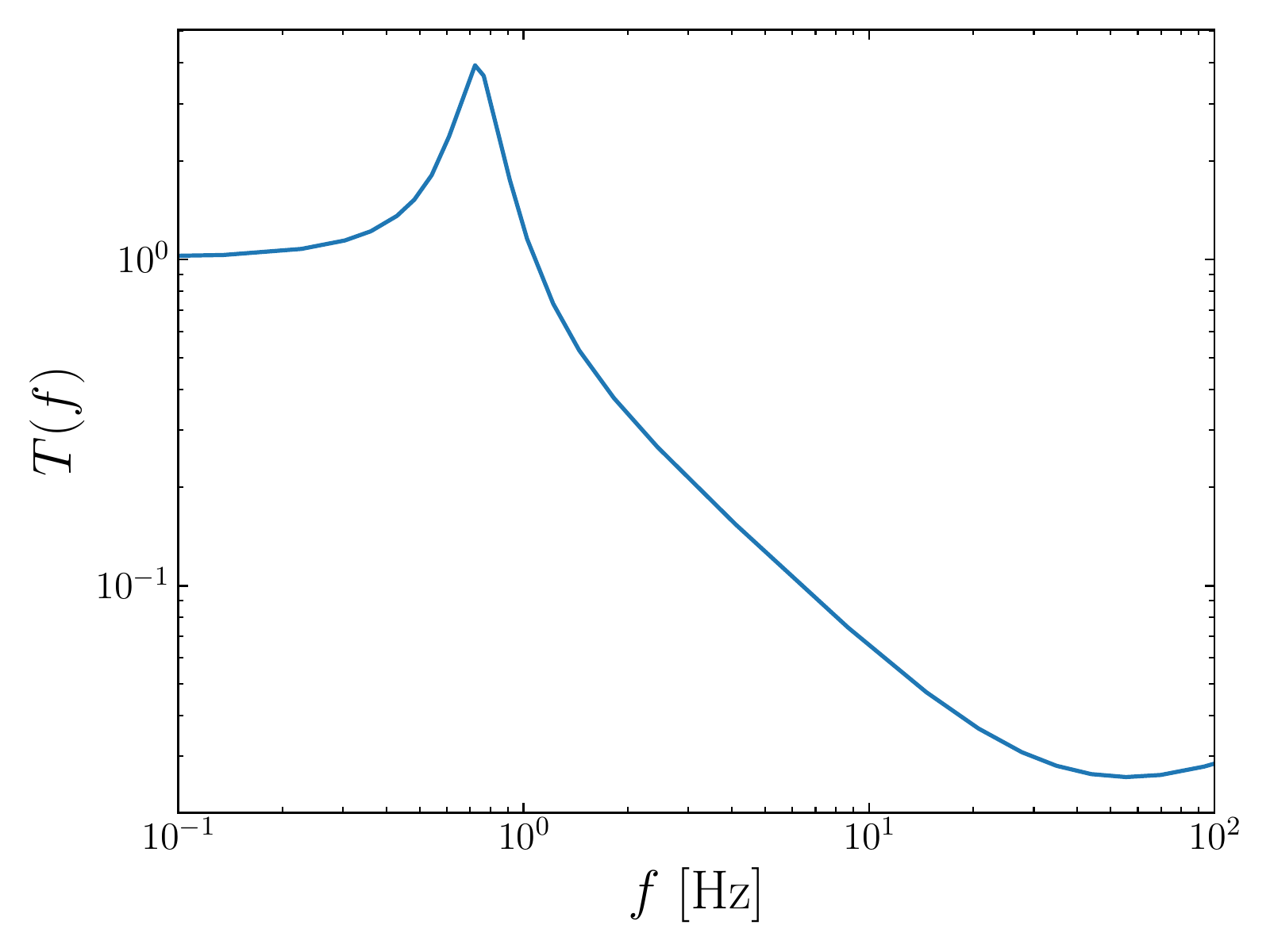} 
\caption{\small Transfer function $T(f)$ vs frequency $f$ for the quadrupole stabilisation system~\cite{quadrupole-stabilisation}.}
\label{f:quad-stab-tf}
\end{figure}

\textbf{The quadrupole stabilisation system} is described in~\cite{quadrupole-stabilisation}. This is an active system which reduces the quadrupole motion. The impact of the quadrupole stabilisation system is included with the transfer function shown in Fig.\,\ref{f:quad-stab-tf}. The quadrupole stabilisation system is effective for suppressing high-frequency ground motion, above 10\,Hz. Low frequency, long wavelength motion is not harmful to the beam. Therefore, the quadrupole stabilisation system was designed to have a transfer function of unity for low frequencies.

\begin{figure}[!htb]
\centering
\includegraphics[width=0.55\textwidth]{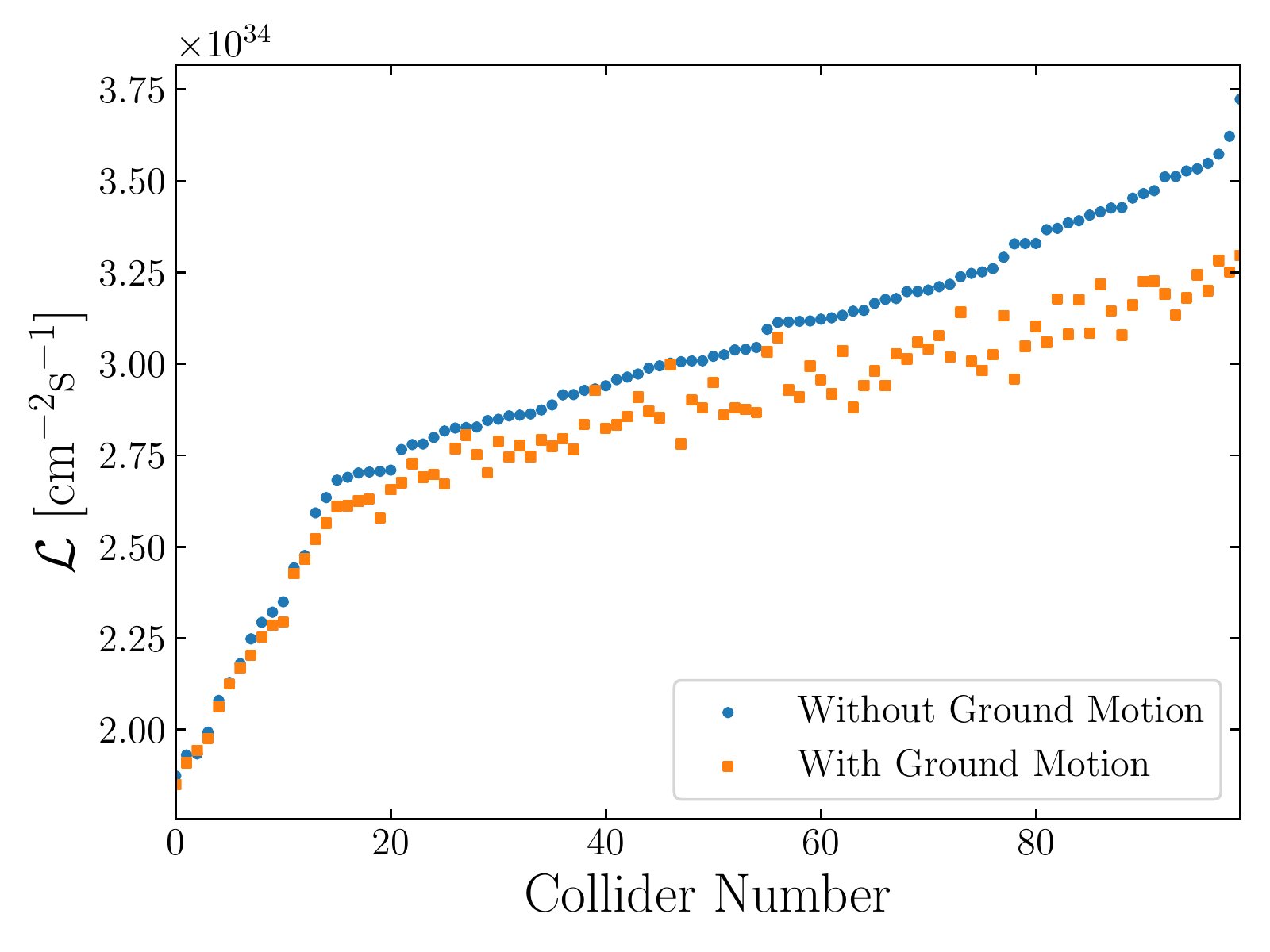}
\caption{\small Luminosity $\mathcal{L}$ vs collider number for 100 tuned colliders with static imperfections: with ground motion (orange square) and without ground motion (blue circle). Collider are ordered in ascending luminosity using the colliders without ground motion.}
\label{f:gm-lumi}
\end{figure}

\subsubsection{Luminosity Loss}
Ground motion model D was simulated with 100 tuned colliders with static imperfections. The impact of the beam-based feedback system and quadrupole stabilisation system was included in these simulations. Figure\,\ref{f:gm-lumi} shows the luminosity of the 100 colliders. The mean luminosity and its standard deviation is
\begin{equation}
\mathcal{L} = (2.8\pm0.3)\times10^{34}~\text{cm}^{-2}\text{s}^{-1}.
\end{equation}
This is a luminosity loss of approximately $0.2\times10^{34}~\text{cm}^{-2}\text{s}^{-1}$ compared to the mean luminosity achieved with static imperfections (Eq.\,\eqref{e:mean-lumi-static}).

Figure~\ref{f:gm-lumi} shows colliders with a higher luminosity suffer from larger luminosity losses due to ground motion. This reflects the fact that colliders with higher luminosities have smaller IP beam sizes and a larger disruption, which makes them more sensitive to luminosity loss due to a beam-beam offset. A luminosity above $2.3\times10^{34}\,\text{cm}^{-2}\text{s}^{-1}$ is achieved by 90\% of colliders with ground motion.

\subsection{Stray Magnetic Fields}
Stray magnetic fields are external dynamic magnetic fields experienced by the beam. Tolerances for stray fields in CLIC at 380\,GeV are presented in~\cite{sf-tolerances, sf-impact}. Stray fields can be divided into three classifications:
\begin{itemize}
\item Natural: stray fields from non-man-made sources. E.g. the Earth's magnetic field.
\item Environmental: stray fields from man-made objects that are not elements of CLIC. E.g. the electrical grid (power lines, sub-power stations) or transport infrastructure (trains, trams, cars, etc.).
\item Technical: stray fields from elements of CLIC.
\end{itemize}
Measurements of each type are described in~\cite{sf-impact, sf-measurements, edu}.

Unfortunately, no realistic model exists for stray fields from technical sources. This is because the spatial distribution of technical sources is a priori unknown. However, stray fields from natural sources exhibit a coherent variation across the length scale of CLIC. Such stray fields can be modelled with a fixed spatial profile across the entire beamline.

\subsubsection{Geomagnetic Storms}
Stray fields from natural sources are discussed in~\cite{balazs-paper}. One of the worst case natural stray field is from a geomagnetic storm, which arises from an interaction of the Earth's magnetic field and solar wind~\cite{balazs-paper}.

\begin{figure}[!htb]
\centering
\includegraphics[width=0.55\textwidth]{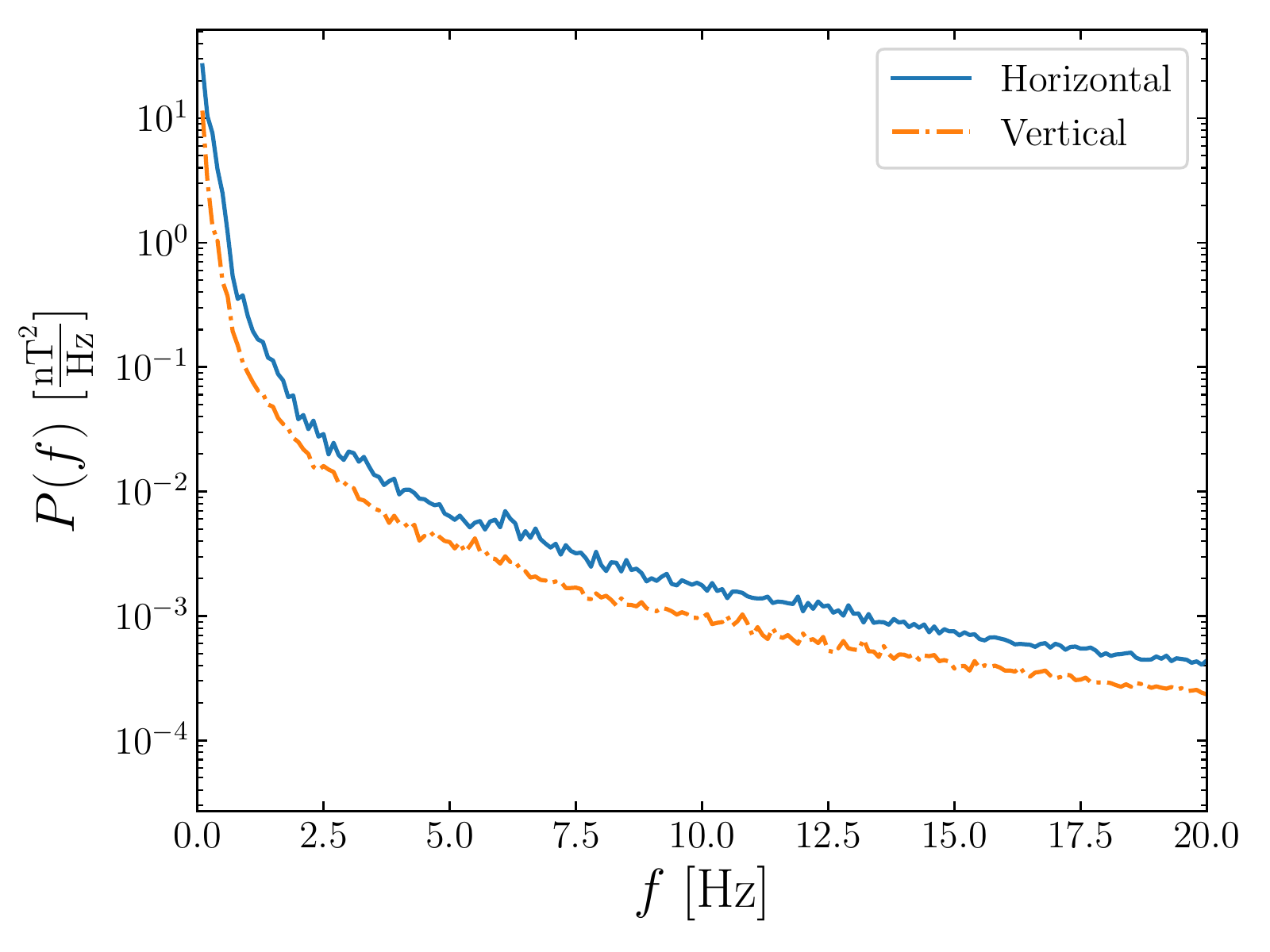} 
\caption{\small PSD $P(f)$ vs frequency $f$ of the stray field during a geomagnetic storm in the horizontal (solid blue) and vertical (dot-dashed orange) direction with respect to the beam.}
\label{f:geomagnetic-storm-psd}
\end{figure}

A representative geomagnetic storm was measured on the 8th June, 2014 in Tihany, Hungary~\cite{balazs}. This location has a similar magnetic environment to CERN. The orientation of the sensor and geometry of CLIC were used to calculate the component of the stray field in the horizontal and vertical direction with respect to the beam. The PSD of the stray field in each direction is shown in Fig.\,\ref{f:geomagnetic-storm-psd}. There is one broadband peak at  a low frequency.

A stray field can be impeded by a beam pipe. A reasonable model for a CLIC beam pipe is a steel cylinder with a 1\,cm inner radius, 1\,mm thickness and a 10-100\,$\mu$m inner copper coating. High-frequency stray fields can induce eddy currents in the beam pipe, which will generate magnetic fields that oppose the external field, thus shielding the beam. However, a 10-100\,$\mu$m copper coating will only be effective at shielding frequencies in the kHz range. As stray fields from geomagnetic storms are at much lower frequencies, the beam pipe will not prevent the stray field from reaching the beam.

\begin{figure}[!htb]
\centering
\includegraphics[width=0.55\textwidth]{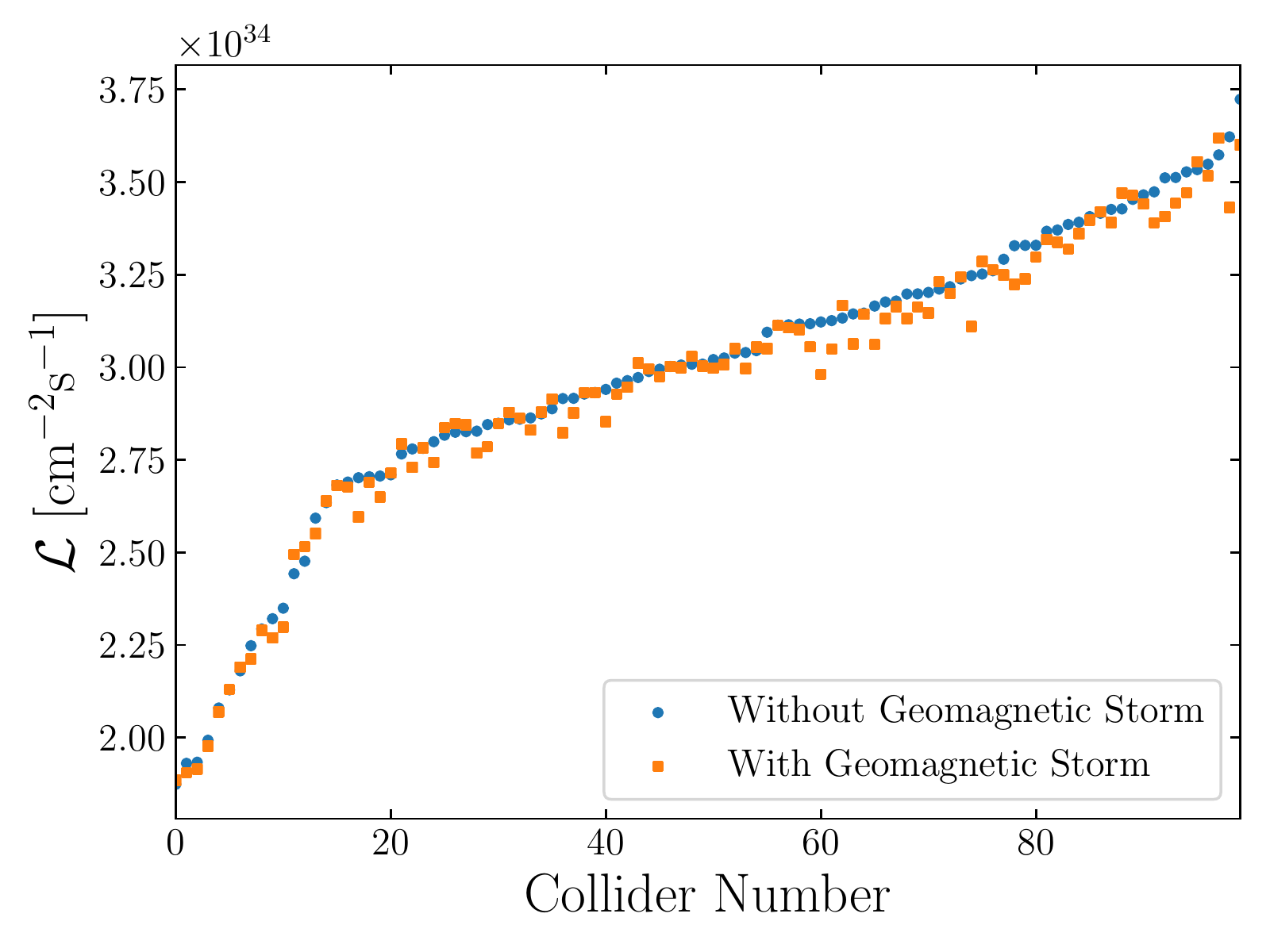}
\caption{\small Luminosity $\mathcal{L}$ vs collider number for 100 tuned colliders with static imperfections: with a geomagnetic storm (orange square) and without a geomagnetic storm (blue circle). Collider are ordered in ascending luminosity using the colliders without a geomagnetic storm.}
\label{f:sf-lumi}
\end{figure}

A geomagnetic storm was simulated in 100 tuned colliders with static imperfections. The direction of the beam (as described by Fig.\,\ref{f:clic-geometry}) was taken into account when calculating the kick from the stray field. The impact of the beam-based feedback system was included. Figure~\ref{f:sf-lumi} shows the luminosity of the 100 colliders. The mean luminosity and 90\% threshold is virtually unaltered by the geomagnetic storm. This is because the impact of the geomagnetic storm can be effectively corrected with the beam-based feedback system. 

\section{Next Steps} \label{s:next-steps}
Performing integrated simulations has been a major step forward to realistically study the beam dynamics in CLIC. Symmetric beamlines for the electron and positron beams were simulated in this work. This is a simplification because the electron and positron beams have different RTMLs. A future step is to include the correct RTML in simulations to track the positron beam.

The simulations in Sec.\,\ref{s:static-imperfections} show that there is an effective tuning procedure that can mitigate the impact of static imperfections. The reliability and accuracy of the luminosity prediction with static imperfections can be improved by simulating more realistic models with less approximations. Furthermore, tuning studies should be performed in the presence of dynamic errors and with realistic signals.

The luminosity estimates with static imperfections are conservative because each beamline was tuned independently. The luminosity can be improved by performing two-beam simulations, which tune the final-focus system to maximise the luminosity for a particular beamline pair.

In this paper, we performed single-bunch simulations including short-range wakefields in the accelerating cavities. The impact of resistive wall wakefields should also be included in the simulation. The impact of long-range wakefields leads to bunches in the train having different offsets. The mean offset of a train is optimised during tuning, which means a relative offset between colliding bunches is possible. Long-range wakefields must be simulated to examine their impact on luminosity. 

In this paper, we have focused on the impact of dynamic imperfections on the short-term luminosity stability. However, there are also processes that impact the long-term luminosity stability. The most important long-term dynamic imperfection is the drift of accelerator elements due to slow ground motion, which is usually modelled using the ATL law~\cite{atl}. Usually, luminosity loss from ATL motion can be fully recovered by performing beam-based alignment. The impact of ATL motion in the ML and the final-focus system has been studied in~\cite{neven} and ~\cite{bds-tuning} respectively. The long-term stability of the collider should be a future study.

\section{Summary} \label{s:summary}
The nominal luminosity target for CLIC is $1.5\times10^{34}~\text{cm}^{-2}\text{s}^{-1}$. In the case without imperfections, integrated simulations show that a luminosity of $4.3\times10^{34}~\text{cm}^{-2}\text{s}^{-1}$ can be achieved. This is almost three times the nominal luminosity target.

Implementing static imperfections and simulating beam-based tuning, integrated simulations of 100 colliders show that an average luminosity of $3\times10^{34}~\text{cm}^{-2}\text{s}^{-1}$ can be achieved, which is twice the nominal luminosity target. 90\% of colliders achieve a luminosity above $2.35\times10^{34}~\text{cm}^{-2}\text{s}^{-1}$. Expressed as a percentage of the nominal luminosity target this is 157\%. Therefore, there is a margin of up to 57\% or greater for dynamic imperfections. This surplus also gives a margin for unforeseen processes which may impact the luminosity.

At the 90\% threshold the luminosity with static imperfections and ground motion is $2.3\times10^{34}~\text{cm}^{-2}\text{s}^{-1}$, which expressed as a percentage of the nominal luminosity target is 153\%. The luminosity loss from stray magnetic fields is negligible. If other dynamic effects such as beam jitter, RF phase errors, etc. are kept within their tolerance, their impact will be on the percent level. Therefore, a significant luminosity surplus of approximately 50\% or greater can be expected for CLIC. A luminosity of $2.8\times10^{34}~\text{cm}^{-2}\text{s}^{-1}$ is achieved by the average collider including static imperfections and ground motion. Expressed as a percentage of the nominal luminosity target, this is 187\%, which is almost twice the target.


\begin{thebibliography}{20}
\bibitem{strategy2013} T. Nakada, CERN Report No. CERN-ESU-003, 2013.
\bibitem{clic-accelerator} A. Robson, P.\,N. Burrows, N. Catalan Lasheras, L. Linssen, M. Petric, D. Schulte, E. Sicking, S. Stapnes, and W. Wuensch, arXiv:1812.07987.
\bibitem{clic-physics} P. Roloff, R. Franceschini, U. Schnoor and A. Wulzer, arXiv:1812.07986.
\bibitem{clic-summary} P.\,N. Burrows, N. Catalan Lasheras, L. Linssen, M. Petri\v{c}, A. Robson, D. Schulte, E. Sicking, and S. Stapnes, CERN Report No. CERN-2018-005-M, 2018.
\bibitem{clic-pip} M. Aicheler, P.\,N. Burrows, N. Catalan Lasheras, R. Corsini, M. Draper, J. Osborne, D. Schulte, S. Stapnes, and M.\,J. Stuart, CERN Report No. CERN-2018-010-M, 2018.
\bibitem{clic-cdr} M. Aicheler, P.\,N. Burrows, M. Draper, T. Garvey, P. Lebrun, K. Peach, N. Phinney, H. Schmickler, D. Schulte, and N. Toge, CERN Report No. CERN-2012-007, 2012.
\bibitem{clic-higgs} H. Abramowicz, et al., Euro. Phys. J. C \textbf{77}, 475 (2017).
\bibitem{clic-top} H. Abramowicz, et al., J. High Energy Phys. \textbf{2019}, 3 (2019).
\bibitem{fabien} F. Plassard, A. Latina, E. Marin, R. Tom\'{a}s, and P. Bambade, Phys. Rev. ST Accel. Beams \textbf{21} 011002 (2018).
\bibitem{rtml-tuning} Y. Han, A. Latina, L. Ma, and D. Schulte, in \textit{Proceedings of the 8th International Particle Accelerator Conference, Copenhagen, Denmark, 2017}, (JACoW, Geneva, 2017), p. TUPIK099.
\bibitem{ml-tuning} N. Blaskovic Kraljevic, and D. Schulte, CERN Report No. CERN-ACC-2018-0053, 2018.
\bibitem{bds-tuning} J. \"{O}gren, A. Latina, R. Tom\'{a}s, and D. Schulte, Phys. Rev. ST Accel. Beams \textbf{23} 051002 (2020).
\bibitem{neven} N. Blaskovic Kraljevic and D. Schulte, in \textit{Proceedings of the 10th International Particle Accelerator Conference, Melbourne, Australia, 2019}, (JACoW, Geneva, 2019), p. MOPMP017.
\bibitem{ryan} R.\,M. Bodenstein, P.\,N. Burrows, F. Plassard and J. Snuverink, in \textit{Proceedings of the 7th International Particle Accelerator Conference, Busan, Korea, 2016}, (JACoW, Geneva, 2016), p. WEPOR009.
\bibitem{banana3} D. Schulte, CERN Report No. CERN-OPEN-2003-014, 2003.
\bibitem{thesis} C. Gohil, DPhil thesis, University of Oxford, 2020.
\bibitem{placet}  The tracking code PLACET, \texttt{https://clicsw.web.cern.ch/clicsw}.
\bibitem{guinea} D. Schulte, Ph.D. thesis, University of Hamburg, 1997.
\bibitem{rtml-csr} H.\,H. Braun, R. Corsini, L. Groening, F. Zhou, A. Kabel, T. O. Raubenheimer, R. Li, and T. Limberg, Phys. Rev. ST Accel. Beams \textbf{3}, 124402 (2000).
\bibitem{local-chromaticity-correction} P. Raimondi and A. Seryi, Phys. Rev. Lett. \textbf{86}, 3779 (2001).
\bibitem{wakefield-monitors} N. Galindo Munoz, N. Catalan Lasheras, S. Zorzetti, M. Wendt, A. Faus Golfe, and V. Boria Esbert, in \textit{Proceeding of the 4th International Beam Instrumentation Conference, Melbourne, Australia, 2015}, (JACoW, Geneva, 2015), p. TUPB054.
\bibitem{rtml-tuning2} Y. Han, A. Latina, L. Ma and D. Schulte, J. Instrum. \textbf{12}, P06010 (2017).
\bibitem{simplex} W.\,H. Press, W.\,T. Vetterling, and S. Teukolsky, \textit{Numerical recipes 3rd edition: The art of scientific computing} (Cambridge University Press, 2007).
\bibitem{font} R.\,J. Apsimon, D.\,R. Bett, N. Blaskovic Kraljevic, R.\,M. Bodenstein, T. Bromwich, P.\,N. Burrows, G.\,B. Christian, B.\,D. Constance, M.\,R. Davis, C. Perry, and R. Ramjiawan, Phys. Rev. Accel. Beams \textbf{21}, 122802 (2018).
\bibitem{font2} J. Resta-L\'{o}pez, P.\,N. Burrows and G. Christian, J. Instrum. \textbf{5}, P09007 (2010).
\bibitem{dynamic-effect-clic-ml} D. Schulte and R. Tomas, in \textit{Proceedings of the 9th International Particle Accelerator Conference, Vancouver, BC, Canada, 2018}, (JACoW, Geneva, 2018), p. TH6PFP046.
\bibitem{jack} J. Roberts, P. Skowronski, P. N. Burrows, G. B. Christian, R. Corsini, A. Ghigo, F. Marcellini, and C. Perry, Phys. Rev. Accel. Beams \textbf{21}, 011001 (2018).
\bibitem{lhc-design-report} O. S. Br\"{u}ning, P. Collier, P. Lebrun, S. Myers, R. Ostojic, J. Poole, and P. Proudlock, CERN Report No. CERN-2004-003-V-1, 2004.
\bibitem{seryi} A. Seryi and O. Napoly, Phys. Rev. \textbf{53}, 5323 (1996).
\bibitem{slac-measurements} C. Adolphsen, et al., SLAC Report No. SLAC-R-474, 1996.
\bibitem{fermilab-measurements} B. Baklakov, T. Bolshakov, A. Chupyra, A. Erokhin, P. Lebedev, V. Parkhomchuk, Sh. Singatulin, J. Lach, and V. Shiltsev, Phys. Rev. ST Accel. Beams \textbf{1}, 031001 (1998).
\bibitem{cms-measurements} A. Kuzmin, Tech. CERN Report No. EDMS:1027459, 2009.
\bibitem{jurgen} J. Pfingstner, Ph.D. thesis, Vienna University of Technology, 2013.
\bibitem{clic-feedback1} G. Balik, B. Caron, D. Schulte, J. Snuverink, and J. Pfingstner, Nucl. Instrum. Methods A \textbf{700}, 163-170 (2013).
\bibitem{clic-feedback2} J. Pfingstner, J. Snuverink, and D. Schulte, Nucl. Instrum. Methods A \textbf{703} 168-176 (2013).
\bibitem{quadrupole-stabilisation} C. Collette, K. Artoos, A. Kuzmin, S. Janssens, M. Sylte, M. Guinchard, and C. Hauviller, Nucl. Instrum. Methods A \textbf{621}, 71-78 (2010).
\bibitem{sf-tolerances} C. Gohil, D. Schulte, and P.\,N. Burrows, CERN Report No. CERN-ACC-2018-0052, 2018.
\bibitem{sf-impact} C. Gohil, M.\,C.\,L. Buzio, E. Marin, D. Schulte, and P.\,N. Burrows, in \textit{Proceedings of the 9th International Particle Accelerator Conference, Vancouver, BC, Canada, 2018}, (JACoW, Geneva, 2018), p. THPAF047.
\bibitem{sf-measurements} C. Gohil, N. Blaskovic Kraljevic, P.\,N. Burrows, B. Heilig, and D. Schulte, in \textit{Proceedings of the 10th International Particle Accelerator Conference, Melbourne, Australia, 2019}, (JACoW, Geneva, 2019), p. MOPGW081.
\bibitem{edu} E. Marin, D. Schulte, and B. Heilig, in \textit{Proceedings of the 8th International Particle Accelerator Conference, Copenhagen, Denmark, 2017}, (JACoW, Geneva, 2017), p. MOPIK077.
\bibitem{balazs-paper} B. Heilig, C. Beggan, and J. Lichtenberger, CERN Report No. CERN-ACC-2018-0033, 2018.
\bibitem{balazs} B. Heilig (private communication).
\bibitem{atl} V.\,D. Shiltsev, in \textit{Proceedings of the 4th International Workshop on Accelerator Alignment, KEK, Tsukuba, Japan, 1995}, (KEK, Tsukuba, 1995), p. 352-381.
\end{thebibliography}
\end{document}